# Spectral evidence from global geomagnetic calibration for stellar-system resonances constraining both quantum physics and geophysics

Mensur Omerbashich, omerbashich@geosciences.online, editor@geophysicsjournal.com, hm@royalfamily.ba

In 1984, Clarke expanded the domain of observed quantum phenomena from the microscopic regimes of atoms and electrons (~$10^1$–$10^3$ particles) to macroscopic superconducting circuits, which he showed exhibit quantum coherence (~$10^9$–$10^{12}$ particles). I demonstrate that coherent quantum behavior persists at vastly larger scales, in Earth's (~$10^{51}$ particles) resonant geophysical cycles, which manifest astro-macroscopic time-crystalline behavior. The arguably most accurate and precise spectral analysis of the conventionally most reliable global geomagnetic calibration (CKGPTS95) revealed that macroscopic and microscopic phenomena are interconnected through cosmos-permeating gravitational resonance networks, thus annulling conventional assumptions of quantum invariance. The analysis identified a planet-dominating 9.35-My fundamental cycle, arising from Earth's orbital and stellar gravitational influences, and resonantly governing geomagnetic reversals, planetary growth, stratigraphic anomalies mimicking mass extinctions, as well as the Great Unconformity. The resonances exhibit non-geophysical features of quantum coherence classically confined to microscopic systems, such as time crystals: discrete time-translation symmetry, fractional harmonic locking, and many-body entrainment. An integrative, high-resolution (10-yr-) view of paleodata, celestial mechanics, and quantum physics reveals that stellar-system resonances impose hierarchical constraints on planetary processes and regulate fundamental quantum properties. Given the ubiquity of tidal phenomena, a resonance-based framework exists in which large-scale celestial dynamics calibrate quantum analogously to how extragalactic dynamics calibrate stellar—thereby constraining particle masses, coupling constants, and universal parameters. This data-driven proof completes my 2006 theoretical derivation of G (and thus gravity) from the speed of light at both quantum and everyday scales, and confirms the Hyperresonance Unifying Theory, which unified those domains using only high-school algebra (https://arxiv.org/abs/physics/0608026, https://arxiv.org/abs/0801.0876).

## Introduction

Quantum mechanics stands as one of the most successful and deeply rooted pillars of modern physics. It describes a universe where particles—quarks, leptons, bosons—exhibit intrinsic properties such as mass, charge, and spin, governed by universal constants that define the so-called fabric of reality. The apparent fine-tuning of these constants—sometimes dubbed the *hierarchy problem*—remains one of the deepest puzzles of physics [1–4]. Yet, despite its successes, the origins and precise values of these fundamental constants remain mysterious. Why is the mass of the electron what it is? What sets the Higgs field's vacuum expectation value? Why are particle masses seemingly fine-tuned, and how do these relate to cosmic parameters?

On the other hand, the Earth's global geological record preserves long-term cyclical signatures—mass extinction events, geomagnetic reversals, sedimentary cycles—that appear to be modulated by celestial influences; periodic behavior in these records has been noted since the earliest works on orbital forcing and extinction periodicity [5]. Classical explanations, such as Milankovitch cycles tied to Earth's orbital eccentricity, obliquity, and precession, account for glacial-interglacial periods that span hundreds of thousands of years. However, the apparent periodicities in mass extinctions, stratigraphic anomalies, and geomagnetic reversals—all spanning millions of years—cannot be fully explained by those models alone. Recent reviews have highlighted the limitations of astrochronological tuning and its tendency to mask long-period resonances. [6]

Furthermore, recent advances challenge the traditional geophysical frameworks by revealing a complex, resonance-driven structure that underpins Earth's long-term dynamics. The interdisciplinary confluence of paleontology, geophysics, and celestial mechanics suggests that Earth's internal processes are enmeshed within a hierarchical resonance network that extends across stellar systems and beyond. This network influences planetary geodynamics and, much like galactic dynamics does to stellar, inevitably calibrates quantum properties at the most fundamental level on Earth and in its vicinity.

The present study posits that the Earth's long-term geophysical and geomagnetic periodicities are governed by gravitational and resonance effects that originate from broader star-system dynamics. Specifically, the ~9.35-My cycle, driven by celestial gravitational interactions, acts as the driver of planetary resonances that not only modulate Earth's internal processes—magnetic field reversals, stratigraphy, and tectonics—but also impose hierarchical constraints onto magnetism and quantum physical constants as measured on Earth and in its vicinity. In that hierarchical, resonance-based universe, large-scale celestial motions serve as the calibration mechanism for both geophysical phenomena and quantum properties, revealing a profound interconnectedness. Such a framework suggests that what we observe as quantum invariants are not absolute but are, at least partially (within the solar system), constrained or determined by macroscopic gravitational resonance networks.

## Methods

The present study applies Shannon theory-based Gauss–Vaniček spectral analysis (GVSA) [7] as a statistically robust and arguably the highest-resolution spectral method to comprehensive global geomagnetic, paleontological, and impact-cratering data. Methodology is in the Supplemental Material.

## Results

*Identification of the 9.35 My Cycle – spectral evidence of planetary resonances*

The GVSA spectra of (the calibration points of) the CKGPTS95 geomagnetic polarity timescale [8] [Supplemental Table 1] reveal a consistent spectral peak at ~9.35 My and its stable foundation, Figs. 1 and 2. The peak is statistically significant across multiple independent datasets, including magnetic polarity reversal records, stratigraphic time-series, and impact signatures.

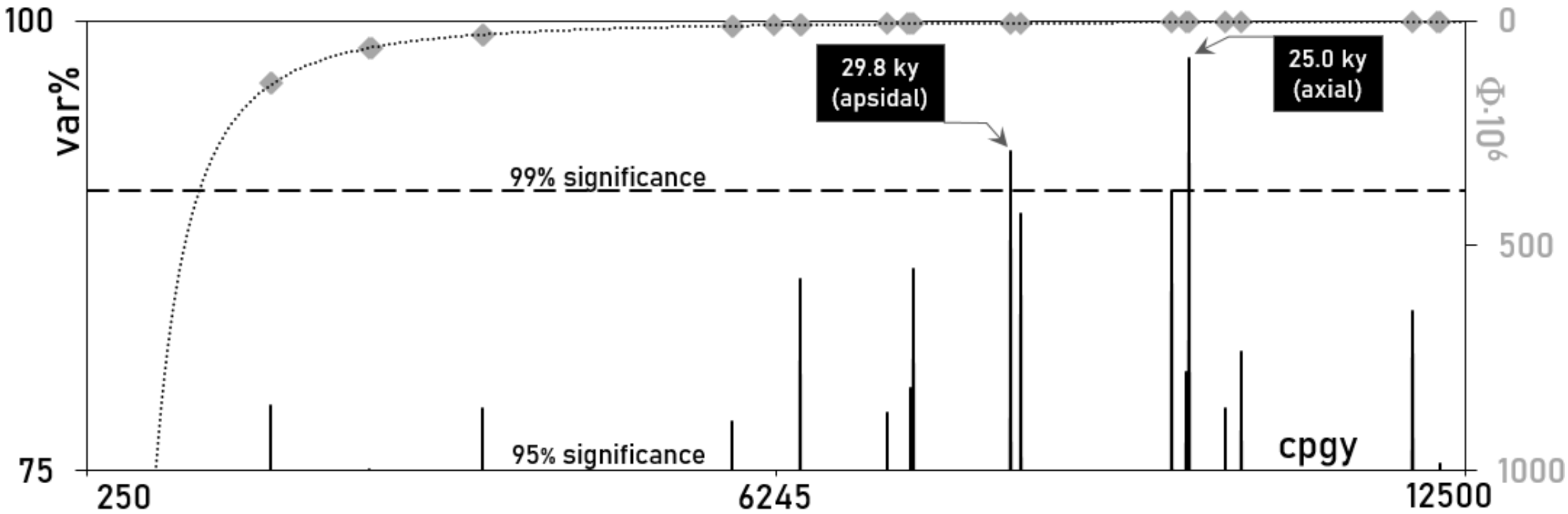

Fig. 1. Gauss–Vaníček spectral analysis (GVSA) [7] of geomagnetic polarity timescale CKGPTS95 [8] reveals ≥99%-significant axial (25 ky) and apsidal (29.8 ky) precessional peaks, confirming a stable foundation for the 9.35-My modulation, Fig. 2.

This spectral maximum coincides remarkably well with Earth's orbital precession period (~25–30 ky)—a cycle primarily modulated by lunar gravitational influences but also affected by planetary perturbations from Venus, Mars, and other bodies, and found across global geological datasets [6]. The phase relationships between diverse datasets exhibit fractional locking—analogous to phenomena observed in quantum time crystals—where stable, discrete periodicities manifest over geological timescales.

Fig. 2 illustrates the spectral peak and harmonic series centered around P≈9.35 My, revealing fractional harmonic relationships $P_i=P/i$ (where P=9.347 My). These fractional harmonics, which include subharmonics (e.g., P/2, P/3), suggest a form of many-body entrainment among Earth's orbital parameters, lunar librations, and planetary alignments, which are collectively amplifying and stabilizing the overall resonance.

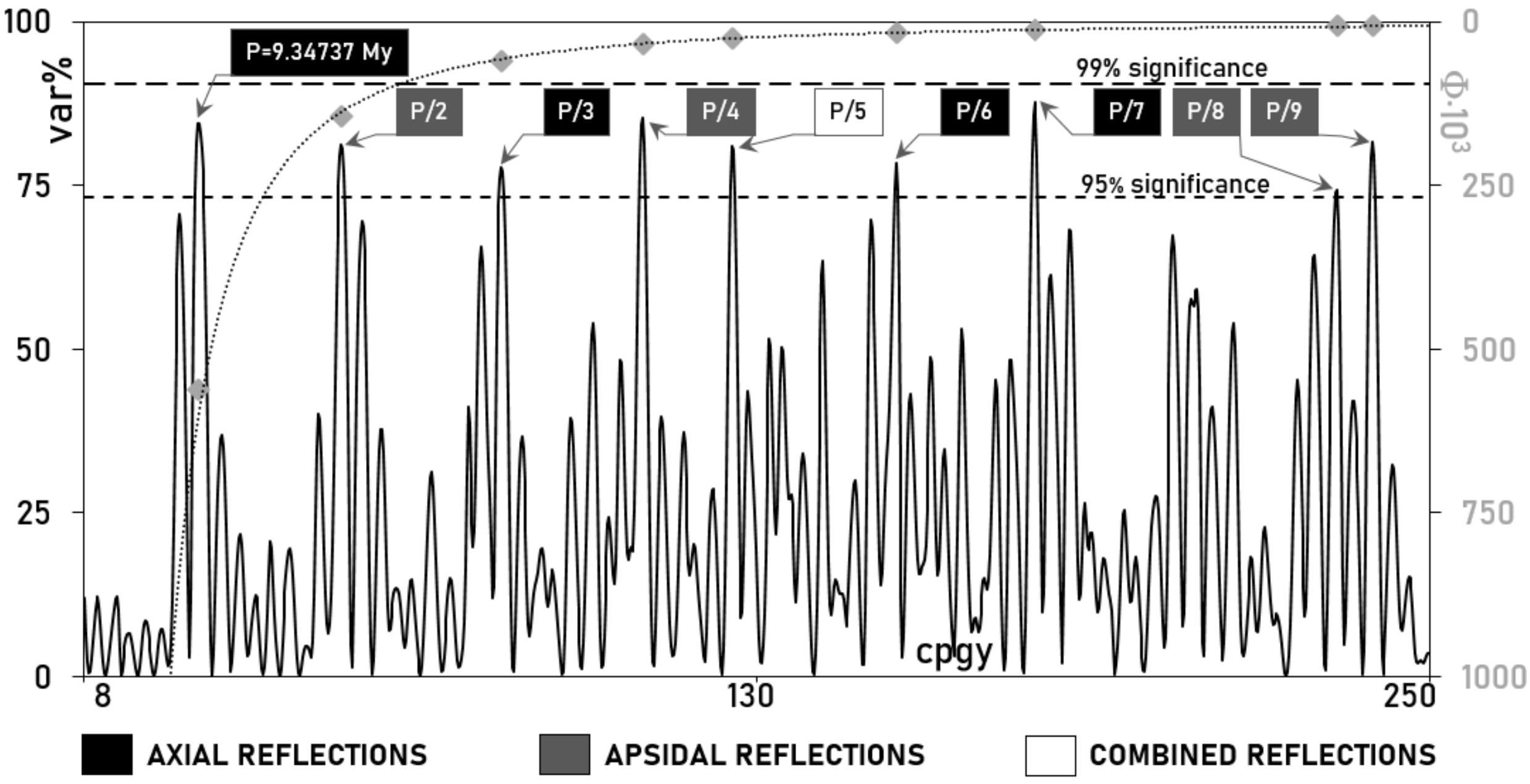

Fig. 2. GVSA spectrum of the CKGPTS95 calibration (1–40-My band) showing the 9.35-My peak and its fractional harmonics $P_i=P/i$, indicative of many-body entrainment.

See Supplemental Material for data sources, statistical tests, and extended spectra.

### *Many-Body Entrainment Framework*

The harmonic series exhibits a network of fractional components—$P_i=P/i$ [Supplemental Table 2]—indicating complex mode locking across multiple planetary and lunar cycles. This many-body entrainment results in a resonance hub that sustains and modulates Earth's internal dynamics. Analogous to coupled oscillators in physical systems, this network could heighten the influence of external gravitational perturbations on Earth’s core processes, mantle convection, and crustal movements.

Supplemental Table 2 lists all periods significant at ≥95% in the GVSA spectrum of CKGPTS95 calibration (1–40 My band). All are harmonics of the 9.347 My resonance, consistent with fractional locking. Full methodological details and additional figures are in the here included Supplemental Material.

Furthermore, the presence of subharmonic locking—a hallmark of quantum time crystals—implies that Earth's long-term geophysical signals do not just reflect simple periodicity but embody fractional, stable phase relationships reminiscent of discrete time-translation symmetry [9–10] found in quantum condensed matter systems.

*Geophysical Manifestations*

The 9.35-My resonance appears to influence multiple geophysical manifestations [11]:

- **Geomagnetic Reversals:** The statistical clustering of polarity reversals around this cycle, with phases of high and low reversal likelihood, suggests a resonance-driven modulation of Earth's hypothetical dynamo. Geomagnetic excursions and rapid reversals could be phase-locked events, temporally modulated by the overarching celestial resonance.
- **Mass Extinctions and Stratigraphic Anomalies:** Paleontological and stratigraphic records display signatures that mimic periodic mass extinctions and unconformities, revealing the 9.35-My cycle. These signatures could be triggered by resonance-induced tectonics and amplified by variations in mantle dynamics, with an impact on volcanic activity and climate.
- **Sedimentary and Stratigraphic Cycles:** The recurring stratigraphic features, including the Great Unconformity, demonstrate timescale periodicities consistent with the resonance framework, suggesting an orbital-driven forcing underlying sediment deposition and erosion patterns.
- **Tectonic and Crustal Reconfigurations:** Large-scale tectonic events, such as rifting, mountain-building, and continental drift episodes, often coincide with resonance peaks, implying external gravitational modulation of Earth's lithospheric processes.

*Quantum-Scale Constraints*

Remarkably, these macro-scale resonance phenomena exhibit features characteristic of **quantum coherence**, including [12–14]:

- **Discrete Time-Translation Symmetry:** Just as quantum time crystals display stable periodicities breaking continuous time-translation symmetry, Earth's long-term records reveal stable, fractional periodicities that endure over millions of years.
- **Fractional Harmonics and Locking:** The presence of harmonic substructures akin to fractional Lock-in phenomena indicates that the Earth's geophysical signals are non-arbitrary, i.e., locked instead into stable fractional modes due to external resonance networks.
- **Many-Body Entrainment:** The complex harmonic relationships suggest collective dynamics involving multiple planetary bodies, akin to many-particle quantum systems known to exhibit entanglement and coherence.

The implication is that Earth's geophysical cycles might embody a form of **macroscopic quantum coherence**, enabled and stabilized by gravitational resonance networks.

The observed spectral and phase coherence suggest a hierarchy where celestial gravitational interactions constrain not only planetary processes but also fundamental quantum properties. The key motif here is the notion that constants of nature—particle masses, charge ratios, and coupling constants—are, at least in part, resonantly calibrated by stellar-system dynamics.

Existing models treat quantum constants as invariant and fundamental, derived from high-energy physics and cosmology. Yet, precise measurements of particle masses, the fine-structure constant, and the Higgs vacuum expectation value appear fine-tuned, prompting questions about their origins.

The data-driven hierarchical paradigm suggests:

- **Particle masses and couplings** might be influenced or constrained by resonance conditions within the gravitational network of stellar systems. For example, particle masses could relate to resonance frequencies that, in turn, are associated with stellar gravitational potentials. This setup is analogous to boundary conditions that set eigenvalues in quantum systems.
- **Dark matter and dark energy** properties might also code unto or be constrained by macro-scale stellar and planetary resonance states, implying that they are emergent phenomena, stabilized by hierarchical gravitational dynamics.
- **Quantum coherence phenomena at macroscopic scales**, counterintuitively, may not be confined to microscopic regimes but can extend or manifest at planetary scales under specific resonance conditions, effectively linking quantum and classical realms.

This perspective challenges the universality presupposed in traditional quantum physics. Instead, it proposes a cosmos where quantum constants "emerge" from, or are "tuned" by, gravitational and celestial resonance networks. Then the constants which we measure reflect a hierarchical structure linking the microcosm to the macrocosm via gravitational resonances.

In summary, these findings indicate an interdependent hierarchical resonance network, where:

- **Stellar system dynamics** is the primary driver shaping planetary geophysics and potentially constraining quantum properties.
- **Planetary internal processes** are governed by large-scale resonant phenomena, such as precession and tidal interactions, which manifest as macroscopic time crystals influencing geophysical and geochemical cycles.
- **Surface and atmospheric processes**, including mass transfer, climate variation, and tectonic activity, are modulated through resonance transfer mechanisms linked to the underlying planetary and stellar dynamics.
- **Quantum and sub-quantum phenomena** are effectively confined and modulated by planetary-scale resonances, suggesting that quantum effects, such as entanglement, tunneling, and coherence, might have a macroscopic geophysical manifestation driven by the hierarchy of resonances initiated at the stellar system level. Just as galaxies are known to impose their own dynamics onto neighbors' constituents, so may stars impose their dynamics onto their own immediate neighbors.

This hierarchical resonance network posits that geophysical phenomena, such as geomagnetic reversals, mass extinctions, and stratigraphic signatures, are not exclusively governed by isolated internal processes or stochastic external perturbations, but are instead emergent properties resulting from coherent, multi-scale interactions rooted in the fundamental dynamics of the stellar system.

Specifically, planetary systems—and by extension, their host stars—act as natural laboratories in which quantum-like coherence manifests in the form of phenomena like time crystals and many-body entrainment. These effects propagate downward, influencing Earth's internal constitution and surface conditions, creating a self-organized, resonant system. In its universal generalization, macro- and micro-physical phenomena interconnect intrinsically through a hierarchy of resonances spanning energy levels from above planetary dimensions to below quantum scales.

## Discussion

While quantum phenomena are conventionally observed at microscopic scales, such as individual atoms or electrons ($\sim10^1$–$10^3$ particles), experiments in superconducting circuits have demonstrated coherent quantum behavior in systems containing $\sim10^9$–$10^{12}$ particles [15], showing that quantum coherence can persist at macroscopic scales far beyond atomic dimensions. In the above, I suggested a conceptual analogy for resonant geophysical cycles in the Earth-Moon-Sun system.

Indeed, the present study extended this macroscopic quantum perspective to planetary scales, suggesting that periodicities in Earth's ($\sim10^{51}$ particles) geological and astronomical data can be interpreted as manifestations of macroscopic time-crystalline behavior. The study presents compelling evidence that Earth's geophysical processes, including geomagnetic reversals and seismic activity, exhibit characteristics of a macroscopic quantum system—specifically, a *discrete time crystal*. This conceptual framework integrates geological, astronomical, and quantum principles, proposing a hierarchical resonance network in which stellar-system dynamics serve as the primary regulatory mechanism influencing planetary internal and surface phenomena.

In such a setup, resonantly stabilized interactions produce coherent, system-level oscillations analogous to those observed in macroscopic quantum circuits, albeit at vastly higher particle numbers and longer timescales. Consequently, quantum-physical phenomena and fundamental parameters are effectively specialized to their host star and stellar system, and may therefore vary among stellar systems if resonance architectures differ. Key findings include:

- **Existence of planetary-scale time crystallinity,** where discrete time translation symmetry shows in Earth's geophysical cycles, including geomagnetic polarity reversals and stratigraphic patterns.
- **Quantum-like coherence** in planetary dynamics, analogous to superconducting circuits, supports the hypothesis of macroscopic quantum phenomena at geological scales.
- **Resonant energy transfer mechanisms** provide a physical basis for Milankovitch cycles, thus explaining long-term climate and magnetic phenomena while implying that Earth's internal processes are in sync with astronomical cycles.
- The **hierarchical order of resonances** links stellar system evolution to planetary and sub-planetary phenomena, as cosmic-scale processes constrain quantum properties and vice versa.

This integrated perspective offers a unifying framework that challenges traditional views that compartmentalize quantum physics, geophysics, and astronomy from each other. The recognition of macroscopic quantum coherence in planetary systems opens avenues for novel interpretations of Earth's history, including mass extinctions, magnetic field behavior, and tectonic activity. It likewise reveals a new perspective on studying stellar systems and exoplanets.

## Conclusions

Using the best available data and analysis tools, I confirmed my earlier theoretical derivation of gravity via the speed of light at both quantum and everyday macroscopic scales, and Hyperresonance Unifying Theory[16], which together unified physics of the two scales by utilizing only high-school algebra (https://arxiv.org/abs/physics/0608026, https://arxiv.org/abs/0801.0876). The here presented discovery revealed a mega-paradigm shift: of Earth's global geophysical phenomena becoming appreciated as emergent properties of a macroscopically quantum, self-organized dynamical system, modulated through mechanical resonance at multiple scales (energy levels) in a step-wise fashion rather than classically through principles of a series of mutually largely incompatible continuum paradigms that we (for historical reasons only) call scientific disciplines and subdisciplines.

By inverting the current scientific worldview, in which the frequency space is merely a byproduct and peculiarity of microscales (with everything beyond regarded as burst-driven if not field-confined; but always merely energy-described), the cross-scale gravitational resonance network and its cross-scale domain took over the place of the sole fundamental position of future physics. This absolute inversion of scientific "dogma" dramatically enhances our predictive capabilities concerning scale-invariant dynamics: on Earth, in issues such as geomagnetic reversals, climate shifts, and planetary stability; and universally, in cross-scale considerations, at all scales at once.

# Supplemental Material

for

"Global geomagnetic calibration reveals that stellar systems resonantly constrain both quantum physics and geophysics"

by Mensur Omerbashich, omerbashich@geosciences.online, editor@geophysicsjournal.com, hm@royalfamily.ba

*This document includes extended methodology, data tables, additional figures, and expanded discussion referenced in the main Article.*

Earth's macroscopic system of planetary cycles exhibits quantum-like coherence analogous to phenomena observed in superconducting circuits. Shannon theory-based Gauss-Vaníček spectral analysis (GVSA) of geomagnetic, impact-cratering, and extinction data revealed that the Earth's precession period, $p\approx26\cdot10^3$ yr, and its harmonics, drive many-body resonant entrainment with astronomical forcing. The resulting quasiperiodic structures in paleodata, phase-locked to obliquity, with ¼-lockstep fractional harmonics, reveal conditions reminiscent of a quantum time crystal—discrete time-translation symmetry and fractional phase locking—at planetary scales. Suppressing precessional frequencies in GPTS-95 calibration residuals isolates the 26.5-My Rampino signal as the driver of geomagnetic polarity reversals, supporting the interpretation that planetary-resonant energy transfer underlies the Milankovitch mechanism. Earth's cross-scale coherence demonstrates that planetary precession is a primary driver of magnetic reversals at all scales within the Earth vicinity and potentially of Earth expansion and phenomena like the Great Unconformity. The view that processes akin to those in quantum systems drive Earth's long-term cycles (geomagnetic reversals, climate oscillations, geologic events) revolutionizes our understanding of Earth's internal processes, framing them as part of a larger, universal resonance phenomenon—within a unifying framework connecting geodynamics, astrophysics, and quantum physics—rather than merely classical mechanics. Quantum-like behavior—here exposed as intrinsically variable beyond Earth's tidal domain—emerges in macroscopic systems on astronomical scales, revealing that quantum physics is a set of physical phenomena localized to and driven by the host star and stellar system and therefore likely fundamentally varying across the known universe. The ubiquity and universality of the tidal phenomenon strongly suggest that particle mass estimates, including those of dark matter, dark energy, and the Higgs boson, could be explained by stellar system and planetary-scale geodynamic resonances rather than solely by fundamental quantum properties.

# Highlights

- Macroscopic quantum coherence is demonstrated in planetary dynamics, analogous to superconducting circuits.

- Quantum behavior is tidally localized, varying across planetary systems via resonance constraints.

- The Earth–Moon–Sun triad forms a measure-preserving dynamical system, acting as a natural discrete time crystal.

- Strong computational evidence links macroscopic planetary dynamics to quantum principles.

- Precession resonance drives many-body entrainment, modulating geomagnetic polarity and stratigraphy through discrete time-translation symmetry (including doubling, halving, and tripling).

- The 26.5-My Rampino cycle is identified as the carrier wave of geomagnetic and climatic resets.

- Accumulated annual timing discrepancies (Equation of Time) project secularly onto paleodata as a global imprinting mechanism.

- Resonant energy transfer provides a physical mechanism for the Milankovitch theory, with phase-locking (¼-lockstep) to obliquity triggering resonant coupling.

- Results suggest a quantum-mechanical basis for Earth's long-term, gradual expansion, sustained by persistent precession-driven resonances.

- Shannon-theory-based Gauss–Vaniček spectral analysis (GVSA) models nonlinear global dynamics directly.

**Preprint**

## 1. Introduction

Any data used to monitor changes rely on a timescale to associate correctly times of changes with the respective values in a record of changes (called then a time series). Geochronology develops methods to achieve the correct timing of such changes in geological time series.

Traditionally, one of the fundamental aims of geochronological studies has been drawing up a timescale that embraces the totality of geological events. Thus a timescale must provide as precise data as possible on the absolute age of the formations. The paleontological and lithological methods often allow stratigraphers to establish, for a studied sedimentary series, a very finely calibrated scale on which the numerous degrees are generally marked by characteristic fossils. Obviously, such scales cannot indicate the absolute ages: they remain relative scales. (IUGS, 1967)

Among various ways to construct a timescale for dating records of paleodata both accurately and precisely (including traditional methods: radiometric dating, biochronology, and magnetochronology), the calibrating of paleodata using astrochronology has become the favored method in the 21st century. In astronomically tuned timescales, dates of samples dated with traditional techniques, particularly those with strata series missing, are purposely and usually at their ends re-matched to the nearest astronomical cycles, most commonly of Earth's and planetary precessions. This tune-up makes very long (spanning millions of years, or My) astrochronologically dated records and their calibration points especially, excellent gauges for spectral analyses of paleodata in the quest for periodicity impressed by physical processes in broadest bands of energies within the Earth's vicinity. Calibration points of a geological timescale are the most accurately (traditionally) dated benchmarks that define that timescale.

Besides the above advantages of astronomical tuning, significant drawbacks also became evident, particularly those due to uncritical ways in which the end-users often approach paleodata record analysis; diversity studies are an example. So despite a looming question of whether intensified diversity over the last 100 My is real or reflects sampling bias and other troubles, reports of such a varied diversity, and the associated (allegedly) periodic mass extinctions from paleodata, have persisted for well over two centuries and have even picked pace in the 1980s (Smith, 2007). This hype resulted in a hotly debated topic in which two sides have emerged: proponents and opponents of periodic mass extinctions; see, for example, Bailer-Jones (2009). This seemingly eternal topic itself has coincided in time with the modernization of geological timescales, which became tied to natural astronomical cycles with the primary goal of enabling a more realistic view of the Earth's system dynamics following the first confirmations of the Milankovitch (1941) theory of astronomically forced climate from deep drilling experiments, e.g., by Emiliani (1955).

Such astronomical calibration of paleodata time series based on cyclostratigraphy, i.e., the identifying and counting cycles in strata, has been praised as an achievement of modern geology; see, e.g., Hinnov and Ogg (2007). However, others were critical for creating an analytical bias against research seeking to modify or replace the Milankovitch theory, e.g., Puetz et al. (2016). Also, there is criticism for contradicting biostratigraphy and inconsistencies due to lack of data in the stratigraphic record, particularly for data older than ~40 My, e.g., Tanner and Lucas (2014). Indeed, the applied astrochronology field carries fundamental computational problems like separating the astronomical signal from climatic and stratigraphic noise and understanding how the astronomical signal propagates through the climate system and into the stratigraphic record (Hinnov, 2013). In addition, determining rapid geomagnetic variations like reversals with adequate temporal precision is difficult since astronomical tuning nowadays is precise to better than 20 ky — a span still exceeding reversals' short duration (Valet & Fournier, 2016). However, none of these potential problems identified in the past seems unsurmountable, provided the right choices in data processing and analysis.

The geological record contains multiple instances of perfect coincidence in the dates of geomagnetic polarity reversals, stage boundaries, and sequence boundaries, creating strong evidence that a single mechanism causes different geological events (Ryskin, 2010). Similarly, Earth's orbital cycles appear to pace the correlations and cyclicity observed in the geologic episodes, indicating a largely periodic, coordinated, and intermittently catastrophic geologic record, which is quite different from the views held by most geologists (Rampino et al., 2021).

While quantum phenomena are conventionally observed at microscopic scales, such as individual atoms or electrons ($\sim 10^1$–$10^3$ particles), experiments in superconducting circuits have demonstrated coherent quantum behavior in systems containing $\sim 10^9$–$10^{12}$ particles (Clarke et al., 1984; 1985). These results show that quantum coherence can persist at macroscopic scales far beyond atomic dimensions, suggesting a conceptual analogy for resonant geophysical cycles in the Earth-Moon-Sun system. Indeed, the present study extends this macroscopic quantum perspective to planetary scales, suggesting that periodicities in Earth's ($\sim 10^{51}$ particles) geological and astronomical data can be interpreted as manifestations of macroscopic time-crystalline behavior. In such system, resonantly stabilized interactions produce coherent, system-level oscillations analogous to those observed in macroscopic quantum circuits, albeit at vastly higher particle numbers and longer timescales. Consequently, quantum-physical phenomena are effectively specialized to their host star and stellar system, and thus may vary significantly across the known Universe.

## 2. Explanatory definitions

Various periodic celestial phenomena affect our planet noticeably every day. Earth's orbital *eccentricity* is one such phenomenon, albeit the relatively mildest. Gravitationally most powerful neighbors — Moon and Sun — cause *nutation*, *obliquity*, and two *precessions*. Those three are the most noticeable of such phenomena. Other planets like Jupiter, Mars, and Venus play minor but noticeable roles. The said celestial phenomena, and some others, cause periodic (therefore constant) changes on Earth. That makes them convenient as natural clocks, as they provide **unity motions**. A unity motion is a fundamental measure of dynamical change, and it means that the actors that carry a change always return to where they started. This repeatable behavior makes their affairs predictable, so we can measure when a deformation they caused — therefore itself periodic — completes once.

A **nutation cycle** of 18.6 yr is the time for the actual celestial pole to circumnavigate the mean pole once and is of no particular concern in the present study. The primary or **axial precession cycle** $p \approx 26$ ky is the time for the quasi-conical motion of the mean celestial pole about the ecliptic pole to complete once. The secondary or **apsidal precession cycle** of $p_1 \approx 29$ ky has no dynamic implications in natural spectra other than accompanying the primary precession periodicity. An **obliquity cycle** $p' \approx 41$ ky is the time for the varying Earth axial tilt (angle between rotational and orbital axes) to change once. An **eccentricity cycle** of $p'' \approx 100$ ky is the time for the deformation in the Earth's orbit about the Sun (due to the divergence of orbital trajectory from a perfect circle) to vary once. **Planetary precession cycles** arise due to the effects of planetary perturbations, primarily by Jupiter and Mars and, to a lesser degree, Venus and Mercury, on the motions of Earth and Moon. Milankovitch (1941) posited that the above phenomena affect the climate on Earth during their unity motions.

The above-stated periodicities vary and oscillate with time due to traction by generally unpredictable geophysical and astrophysical forces. Therefore, measuring the actual periodicity of the above astronomical cycles accurately and precisely at any given instant — past, present, or future — requires timing systems of the highest precision. Astronomers and geodesists have developed various reference systems, frames, and means for that purpose. The time systems are **solar time**, based on the Earth's revolution about the Sun, and **sidereal time**, based on the Earth's rotation about its axis. Astrophysical forces affect the former time system mostly and geophysical the latter. While solar time helps us with the calendar and seasons, sidereal time helps measure the Earth's real diurnal ("over the course of one day") rotation. Therefore, given that the Earth's rotation gets affected by the above-mentioned astronomical phenomena while the Sun is not, sidereal time is more relevant in Earth sciences than solar time is. Specifically, due to precession and nutation, Earth's rotation is subject to the motion of the **equinox** — the instant at which the Earth's equatorial plane passes through the center of the Sun plane. Thus sidereal time reflects the actual rotation of the Earth, and we can determine it from observations of stars, artificial satellites, and extragalactic radio sources. Then, there is **apparent sidereal time** that measures Earth's rotation with respect to the true (actual) equinox. Of relevance then here is the **apparent diurnal motion** of the Sun — an effect due to both Earth's non-uniform daily rotation and the orbital motion around the Sun. For each local meridian, there is always a corresponding local sidereal time. Then, **local mean time** is the true time at a given location, and the (always local-only) **apparent solar time** is the apparent motion of the actual Sun based on one apparent solar day as a measure of the time between two successive returns of the Sun to the local meridian. Likewise, one **solar year** (also called a tropical year) measures the time between two consecutive returns of the Sun to the same position in the sky. While defining our calendars and seasons, it differs by 20 min 24 s from one **sidereal year** that measures the time for the Earth to orbit the Sun once relative to fixed stars.

Of all the above-mentioned astronomical periods, apparent times are virtually (still, never absolutely) independent only of precessions. (Seidelmann, 1992) This last point is the motivation behind the minimalistic approach of tying the geological timescales primarily to the precessional cycles. The present study addresses the apparent times problem, their previously neglected imperfections in particular, and the effects of those imperfections on paleodata analyses.

Here *precession resonance* marks orbital resonance, while *precessional resonance* includes energy transfer from an orbital forcing (or disturbances from it) as impressed into solid or gaseous (solidified) matter that had recorded it or otherwise given in to it. The energy transfer also causes classical *mechanical resonance* — the resonating vibration of matter phenomenon, notably in solids, occurring when the natural-vibration period of a body with a certain mass coincides with a vibrational period (or its fractional multiple) of another mass body.

A *subharmonic mechanical resonance* occurs when the matched period represents a subharmonic nonlinear vibration $n/(mT)$; $n \in \aleph$, $n/m \in (0, 1)$; $n > 1 \land n \in \aleph$. In macroscopic physical systems, such as the Solar system and its objects individually, orbitally forced mechanical resonances can give rise to the *Faraday instability* — a phenomenon characterized by polygonal morphology and patterned topographies. See Omerbashich (2020a) for details on mechanical resonances throughout the Solar system.

I term astronomical cycles and respective periods of duration interchangeably throughout, noting that, in spectral analysis, periods are durations; cycles are events repeated over time.

## 3. Methodology

To examine the effects of astronomical forcing on paleodata, I spectrally analyze age calibration of the currently accepted timescale of geomagnetic polarity reversals, spanning the Cenozoic (0–84 My ago) — the revised Geomagnetic Polarity Time Scale 1995 (CKGPTS95; also cited in the literature as GPTS-95) by Cande and Kent (1995), Table 1 and their Table 1. The CKGPTS95 calibration leans on the South Atlantic Anomaly (where the strength of the geomagnetic field is decreasing most rapidly) sequence and consists of nine control points adjusted astrochronologically. At the control points, the sedimentary record is tied in with the astronomical record (Hilgen, 1991) so that the CKGPTS95 has excellent sensitivity primarily to the Earth's (theoretical) 25.7-ky axial precession (Shackleton et al., 1990) and possibly to the ~28-ky (apsidal) precession as well (Hilgen, 1991), while tuning to an eccentricity cycle of 412.9-ky was also performed, which was unlike values other workers used (Puetz et al., 2016). As embedded in the timescale itself, the so-boosted (systematic) signal becomes an integral part of the overall spectral information, thus easing energy-band (variance-) stratification and enabling the separation of actual astronomical periods from any (individually relatively weak) harmonics. Besides, the CKGPTS95 has stood the test of time with its peers — unlike any other timescale.

**Table 1**. Age calibration points for two related geomagnetic polarity timescales extending to end-Campanian. Left panel: for the currently accepted timescale CKGPTS95 (Cande & Kent, 1995). Right panel: for the superseded timescale CKGPTS92 (Cande & Kent, 1992). The facts that calibration points are the most accurate representation of a geological timescale, that CKGPTS95 has stood the test of time like no other timescale, and that the Gauss–Vaníček Spectral Analysis (GVSA) can draw the most accurate spectra from only three values — are used in the present study to investigate relations among astronomical forcing and reflections or harmonics recorded in paleodata.

| CKGPTS95 | | | CKGPTS92 | | |
|---|---|---|---|---|---|
| Polarity Chron | Age [My] | S. Atlantic dist. [km] | Polarity Chron | Age [My] | S. Atlantic dist. [km] |
| C3n.4n(o) | 5.23 | 84.68 | C2An(0.0) | 2.60 | 41.75 |
| C5Bn(y) | 14.80 | 290.17 | CSBn(0.0) | 14.80 | 290.17 |
| C6Cn.2r(y) | 23.80 | 501.55 | C6Cn,2r(0.0) | 23.80 | 501.55 |
| C13r(.14) | 33.70 | 759.49 | C13r(.14) | 33.70 | 759.49 |
| C21n(.33) | 46.80 | 1071.62 | C21n(.33) | 46.80 | 1071.62 |
| C24r(.66) | 55.00 | 1221.20 | C24r(.66) | 55.00 | 1221.20 |
| C29r(.3) | 65.00 | 1364.37 | C29r(.3) | 66.00 | 1364.37 |
| C33n(.15) | 74.50 | 1575.56 | C33n(.15) | 74.50 | 1575.56 |
| C34n(y) | 83.00 | 1862.32 | C34n(0.0) | 83.00 | 1862.32 |

I deal with spectral bands from just above the Earth's ~26 ky precessions down to 40 My since astronomical timescales (cyclostratigraphy-based numerical timescales) are reasonably well established for much of Cenozoic time (from the beginning of the Oligocene (~34 My) to the present); older parts have less-complete disconnected cyclostratigraphies referred to as "floating astrochronologies" (Tanner & Lucas, 2014). Besides, beyond 40 My, the accuracy of astrochronological timescales critically depends on the correctness of orbital models and radio-isotopic dating techniques, e.g., Westerhold et al. (2012; 2015).

To obtain the periodic signal, I use the Gauss–Vaníček rigorous method of spectral analysis (GVSA) by Vaníček (1969, 1971). The GVSA belongs to the least-squares class of spectral analysis techniques, has many advantages over the Fourier class of spectral analysis techniques in analyzing sparse natural data of long spans (Press et al., 2007), and has proven itself by providing absolute extraction accuracy in analyzing even extremely gapped paleodata (extinction) records (Omerbashich, 2021, 2007b, 2006). A GVSA spectrum, $s_j$, is obtained at a spectral resolution $k$ (here 1000 spectral values throughout), for $k$ corresponding periods $T_j$ or frequencies $\omega_j$ and output with spectral magnitudes $M_j$, as:

$$s_j(T_j, M_j);\ j \in \mathbb{Z},\ j = 1 \dots k \ \wedge \ k \in \aleph\,. \tag{1}$$

In its simplest form, i.e., when there is no *a priori* knowledge on data constituents such as datum offsets, linear trends, and instrumental drifts, a GVSA spectrum $\boldsymbol{s}$ is computed as (Omerbashich, 2004):

$$\boldsymbol{s}(\omega_j, M_j) = \frac{\boldsymbol{l}^{\mathrm{T}} \cdot \boldsymbol{p}(\omega_j)}{\boldsymbol{l}^{\mathrm{T}} \cdot \boldsymbol{l}}\,, \tag{2}$$

obtained after two orthogonal projections. First, of the vector of $m$ observations, $\boldsymbol{l}$, onto the manifold $Z\,(\boldsymbol{\Psi})$ spanned by different base functions (columns of $\mathbf{A}$ matrix) at a time instant $t$, $\boldsymbol{\Psi} = [\cos\omega t, \sin\omega t]$, to obtain the best fitting approximant $\boldsymbol{p} = \sum_{i=1}^{m} \hat{c}_i\, \boldsymbol{\Psi}_i$ to $\boldsymbol{l}$ such that the residuals $\hat{\boldsymbol{v}} = \boldsymbol{l} - \boldsymbol{p}$ are minimized in the least-squares sense for $\hat{\boldsymbol{c}} = (\boldsymbol{\Psi}^{\mathrm{T}}\mathbf{C}_l^{-1}\boldsymbol{\Psi})^{-1} \cdot \boldsymbol{\Psi}^{\mathrm{T}}\mathbf{C}_l^{-1}\boldsymbol{l}$. The second projection, of $\boldsymbol{p}$ onto $\boldsymbol{l}$, enables us to obtain the spectral values, Eq. (2). Vectors $\boldsymbol{u}_j = \boldsymbol{\Psi}^{\mathrm{T}}\, \boldsymbol{\Psi}_{\mathrm{NK}+1}$ and $\boldsymbol{v}_j = \boldsymbol{\Psi}^{\mathrm{T}}\, \boldsymbol{\Psi}_{\mathrm{NK}+2}$, $j = 1, 2 \ldots$. $\mathrm{NK} \in \aleph$, compose columns of the matrix $\mathbf{A}_{\mathrm{NK,NK}} = \boldsymbol{\Psi}^{\mathrm{T}}\, \boldsymbol{\Psi}$. Note here that the vectors of known constituents compose matrix $\widehat{\mathbf{A}}_{\mathrm{m,m}} = \widehat{\boldsymbol{\Psi}}^{\mathrm{T}}\, \widehat{\boldsymbol{\Psi}}$, in which case the base functions spanning the manifold $Z(\boldsymbol{\Psi})$ get expanded by known-constituent base functions, $\widehat{\boldsymbol{\Psi}}$, to $\boldsymbol{\Psi} = \left[\widehat{\boldsymbol{\Psi}}, \cos\omega t, \sin\omega t\right]$. For a detailed treatment of GVSA with known data constituents, see Wells et al. (1985). Subsequently, the method got simplified into non-rigorous (strictly non-least-squares) formats like the Lomb-Scargle technique created to lower any computing burden of Vaníček's pioneering development, but that is no longer an issue. At the same time, the conventional Fourier transform and spectrum are just special cases of a more general least-squares formulation (Craymer, 1998).

As mentioned, the GVSA provides total (absolute) accuracy in extracting periods from natural data sets — meaning at the prescribed accuracy of analyzed data themselves — of twice the sampling step or data accuracy (Omerbashich, 2020b; 2020a; 2007a). Fed raw data, GVSA outputs spectral peaks with spectral magnitudes in variance percentages (var%) against linear background noise levels, or dB (Pagiatakis, 1999). Such processing enables relative spectral computations whose results for physical systems are then variance- (directly energy-) stratified. This procedure is unlike that used for any other spectral analysis method. Relative analyses using GVSA include detecting field dynamics (Omerbashich, 2006; 2003), separation of forcers and harmonics, suppressing selected periods to reveal underlying dynamics, and computing spectra of spectra to separate overlying dynamics from the respective system already dynamical. GVSA revolutionizes physics by enabling direct computations of nonlinear dynamics, rendering classical approaches such as spherical approximation obsolete (Omerbashich, 2024b; 2024a; 2023b; 2023a). I use this multifaceted ability of the GVSA to separate harmonics from their drivers and identify oscillation triggers.

GVSA is strict in that, besides estimating a uniform spectrum-wide statistical significance in var% for the desired level, say 95%, in a spectrum from a time series with *m* data values and *q* known constituents as $1–0.95^{2/(m–q–2)}$ (Steeves, 1981) (Wells et al., 1985), it also imposes an additional constraint for determining the validity of each significant peak individually — the fidelity or realism, Φ. *Fidelity* is a general information measure in advanced statistics based on the coordinate-independent cumulative distribution and critical yet previously neglected symmetry considerations (Kinkhabwala, 2013). In communications theory, fidelity measures how undesirable it is (according to some fidelity criterion we devise) to receive one piece of information when another is transmitted (Shannon, 1948). In GVSA, fidelity thus is defined in terms of the theory of spectral analysis as a measure of how undesirable it is for two frequencies to coincide (occupy the same frequency space of a sample). Then a value of GVSA fidelity is obtained as that time interval (in units of the timescale of the time series analyzed) by which the period of a significant spectral peak must be elongated or shortened to be π-phase-shiftable within the length of that time series. As such, Φ measures the unresolvedness between two consecutive significant spectral peaks (those that cannot be π-phase-shifted). When periods of such spectral peaks differ by more than the fidelity value of the former, those peaks are resolvable. As the *degree of a spectral peak's dependence or tendency to cluster*, this criterion reveals whether a spectral peak can share a systematic nature with another spectral peak, e.g., be part of a batch or an underlying dynamical process like resonance or reflection. The spectral peaks that meet this criterion are listed in the LSSA software output among insignificant and the rest among significant (hereafter: *physically-statistically significant spectral peaks* or just (fully) *significant* for short).

Then in the present study, input data are used in their raw form, i.e., without preprocessing such as dataset padding or various types of filtering (including windowing or tapering); historically, the intention for such techniques for vastly massaging data was to help overcome drawbacks of classical methods such as Fourier's. Finally, I apply no post-processing, used by some to enhance spectra. The declared precision of the periodicities extracted in the present study is ±10 years throughout. While the minimum number of values GVSA can compute a frequency spectrum from is three, Table 1, the method does not depend on the Nyquist frequency (Craymer, 1999).

The above factors, combined with the unprecedented abilities of GVSA, such as handling extremely (>99.99%) depopulated records in their raw form with ease and extracting both field dynamics and over-dynamics (dynamics of dynamics), made the CKGPTS95 the most suitable gauge for examining effects that astronomical cycles and their modulations have on paleodata timing methods and consequently on any other natural periodicity in those data as well. Thus, separating those two types of periodicity (in timescale; in paleodata) should be possible using GVSA.

## 4. Probing CKGPTS95 timescale of geomagnetic polarity

The GVSA of the CKGPTS95 nine calibration values, Table 1, has revealed two ≥99%-significant periodicities in the 0.02–40.00-My full band. Those periods are Earth axial precession, p=25.00 ky (theoretical median: 25.77 ky), and Earth apsidal precession, $p_1$=29.81 ky (theoretical maximum: 29.00 ky), Fig. 1. Note here that the p=25.00 ky is the strongest and statistically most significant periodicity found in the present study; Olsen (1986) reported the same axial-precessional value in the large-lakes sedimentary record, and it often is used to tune geological timescales astronomically, thereby methodologically (artificially) boosting its power but not overwhelmingly, i.e., not absorbing other periodicities to suppression below significance. Thus, and as was desired from a conventional quality timescale, there were no further ≥99%-significant periodicities in the 1–40-My narrowed band, which, however, did contain ten peaks significant at ≥95%, Fig. 2. The longest was P=9.34737 My, while the remaining 9 are its reflections ordered in a series $P_i$=P/i, i=1…n; n $\in \mathbb{Z}^+$.

The P=$P_1$ itself is the same peak detected recently by Omerbashich (2021) from the non-marine tetrapods' extinctions record that here is particularly suitable due to its insensitivity to the ocean-tidal component as the most persistent systematic noise constituent. Thus, P turned out to be just a circular (2π-) modulation of the axial-precessional period, p. Namely, from the vector representation of harmonic oscillation (Den Hartog, 1985), for the 1-yr base oscillation in the case of the Earth, one obtains theoretically: $P_{theor}=360°\cdot p_{theor}$=9.27792 My; $p_{theor}$=25.772 ky, which is matched by the here measured P, to within 7‰ or 69.45 ky or ~2.7 × precession (or approximately twice the CKGPTS95 tuning accuracy, of one precession cycle).

This highly-precise computation of a resonant response and its driver (both the enhanced p and naturally arising 2π-phase-shifted p) is yet another demonstration of the ability of the GVSA to extract spectra with the absolute facility (of twice the data accuracy), even from very long-spanning and heavily gapped records of data. Note here that a commonly understood statistical fidelity threshold that indicates a physical process is $\phi$=12 (Omerbashich, 2006). If most of the system periods obtained from sparse data were previously reported or are otherwise known as natural periods (astronomical or physical) or their phase-shifted or time-symmetrical modulations (multiples or fractions), then of interest is relative $\phi$ primarily.

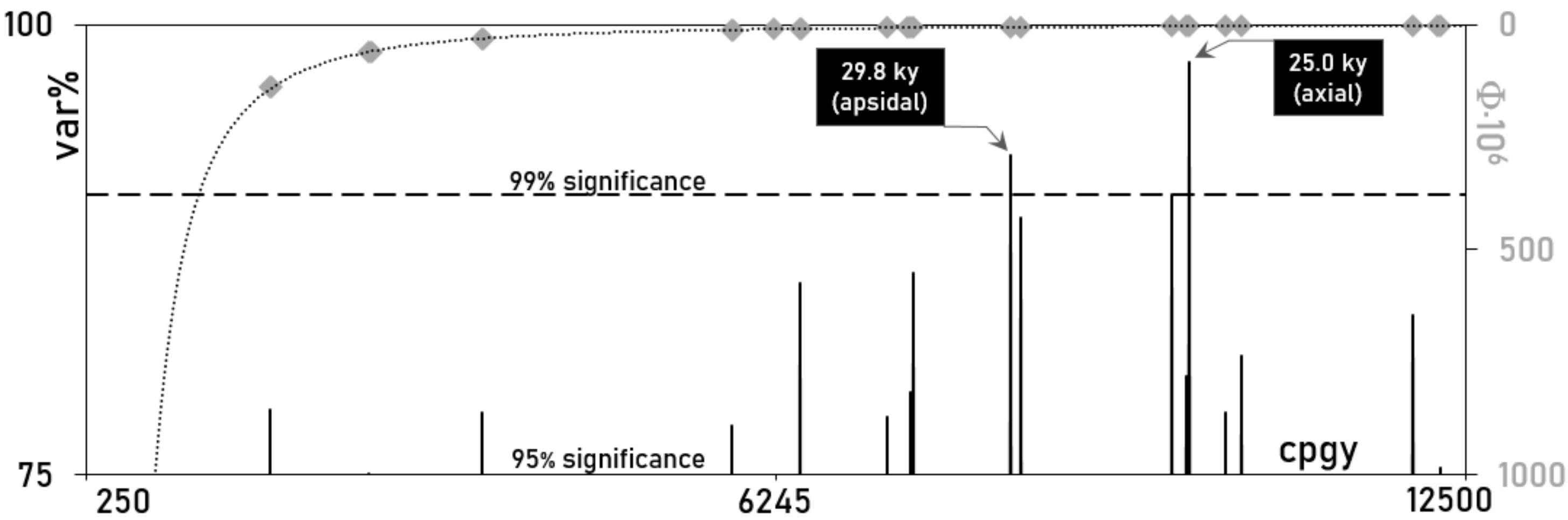

**Figure 1 (replicated from the main Article for completeness)**. Gauss–Vaníček spectrum of the currently accepted geomagnetic polarity timescale CKGPTS95 (calibration), Table 1, in the 20 ky–40-My full band. The only ≥99%-significant periods are the apsidal-precessional, $p_1$=29.81 ky (theoretical maximum: 29.00 ky), and the axial-precessional, p=25.00 ky (theoretical median: 25.77 ky). Spectral peaks' respective statistical fidelity values overlaid with the power trend. Frequencies are depicted (here and throughout) in cycles per galactic year (here 250 My), c.p.g.y.

Mathematically suppressing, see, e.g., Taylor and Hamilton (1972) and Wells et al. (1985), i.e., ignoring (thereinafter: enforcing) the axial-precessional period as realized by CKGPTS95, of 0.02500 My, and apsidal-precessional period, of 0.02981 My, has made them both disappear below the ≥99%-significance level, enabling a separation of the two respective groups of reflections as well, Fig. 2 (shows 1–40 My narrow band, no enforcing). The spectra of the now superseded timescale CKGPTS92, previously created by Cande and Kent (1992), contained in the 0.02–40-My full band the axial-precession period, as 0.025 My, then its 0.02401-My reflection, and the overestimated 0.03863-My apsidal precession period — all three at the ≥99%-significance level and with fidelities negligible at $\phi<10^{-5}$. At the ≥95%-significance level, the superseded GPTS92 timescale was also found periodic in the 1–40 My long band, again with the P=9.34737-My circular precessional modulation, whose fidelity was a still low $\phi$=0.5. However, there detected were five reflections, all disordered: 4.31208 My with $\phi$=0.1, then 4.13793 My with $\phi$=0.1, 1.72606 My with $\phi$=0.02, 1.64831 My with $\phi$=0.02, and 1.40048 My with $\phi$=0.01.

Finally, the GVSA spectrum of the current timescale CKGPTS95 in the 40–80-My extended band contained no peaks at any level of significance. Note that enforcing does not denote a literal data intervention, as no data get removed. Rather systematic contributions (influences) of select periods are suppressed (Wells et al., 1985) via the process of enforcing selected periods. When further systematic information is sought and presumed to underlie an already systematically dominated record of interest, the residual time series (leftovers from the enforcing) can also be spectrally analyzed. This procedure can expose secondary dynamics hidden due to the record's dominant periodicity, like resonance.

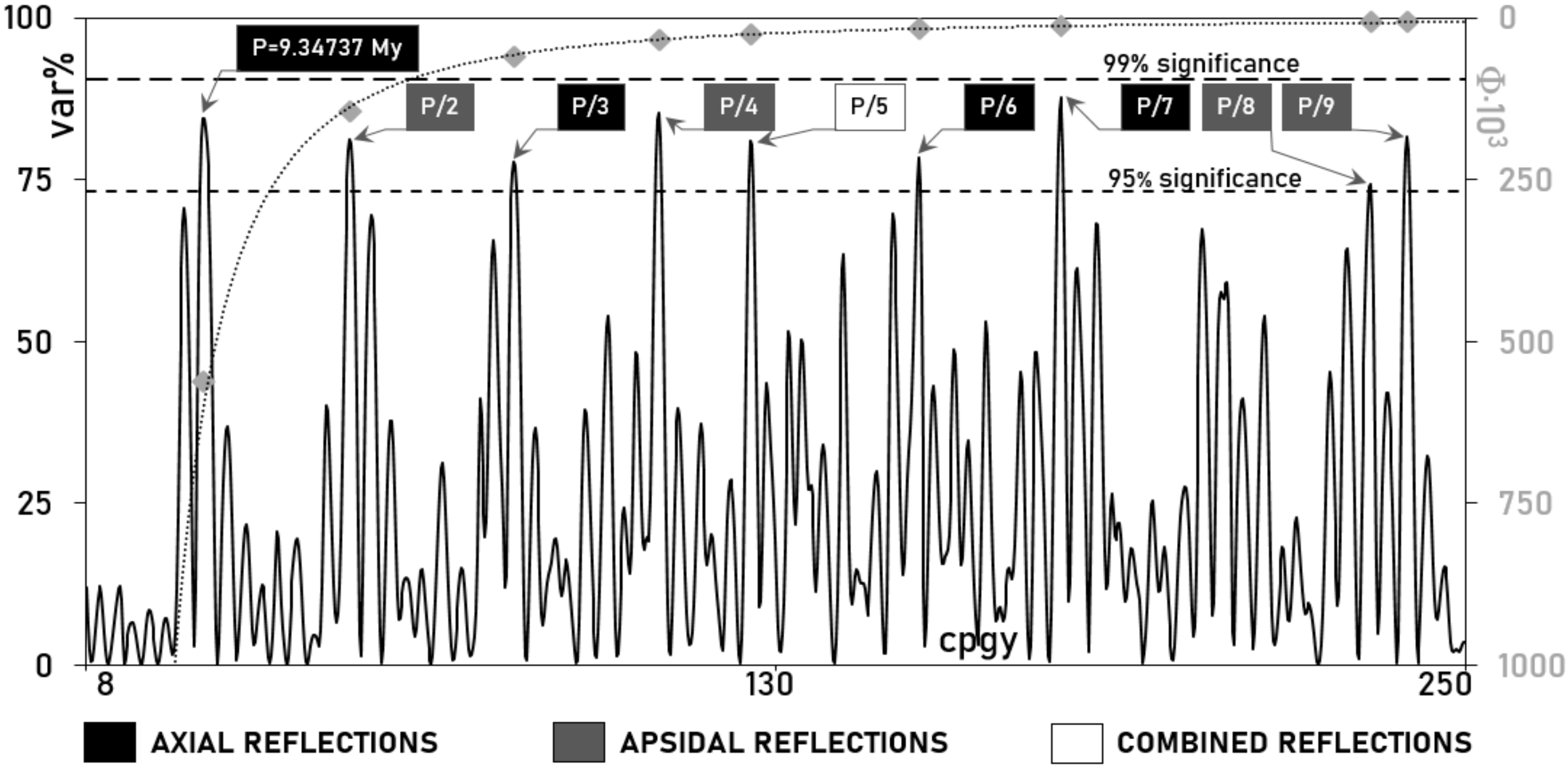

**Figure 2 (replicated from the main Article for completeness)**. The GVSA spectrum of the currently accepted geomagnetic timescale CKGPTS95 (calibration), Table 1, in the 1–40-My long band, from raw data (no preprocessing including any padding, filtering, or tapering). There are no ≥99%-significant spectral peaks, as expected after narrowing the band, which has left out only two ≥99%-significant periods from the spectral band of interest, Fig. 1, as they both are <1 My. At the ≥95%-significance level are the lead (main) modulation P=9.34737 My, and all of its eight fractional harmonics (2π/n modulations of the axial precession), ordered in a series $P_i$=P/i, i=1…n, Table 2. Further analysis, in which axial- and apsidal-precessional periods get mathematically suppressed, e.g., Taylor and Hamilton (1972), i.e., ignored (thereinafter: enforced), has enabled the separation of the reflections, revealing four of nine periods as predominantly axial precession's reflections and four apsidal precession's reflections, respectively. One of the reflections was due to a combined effect of the two precessions. As in Fig. 1, the respective statistical fidelity values for spectral peaks are shown with the power trend overlaid.

**Table 2**. The ≥95%-significant periods, Fig. 2, detected in the 1–40-My-band GVSA spectrum of the CKGPTS95 geological timescale calibration, Table 1. All turned out to be harmonics of the lead period, P=9.34737 My, itself a 2π (circular) modulation of the precession cycle, p=25.772 ky to within 7‰. The precession's circular (annual-secular) modulation often is reported in the literature from paleodata of seemingly unrelated types. The statistical tests show that practically there are no outliers (at 1σ) and that all of the series constituents as extracted belong to the same population, i.e., an underlying physical process. Here, while the detected ensemble (train) of periods represents harmonics of P, those periods individually at the same time represent reflections of p, so the two labels are used interchangeably in the present study, depending on which of the two lead periods' roles needs emphasis.

| GVSA of geological timescale CKGPTS95 calibration, ≥95%-significant periods (1–40-My band) | | | | | | | | |
|---|---|---|---|---|---|---|---|---|
| i | GVSA period $P_i$ [My] | $\Phi_{period}$ | mag [var%] | harmonic | $_{theor}P_i = P / (i+2)$ [My] | $\Delta_i = P_i - {}_{theor}P_i$ [My] | $\Delta_I$ [%] | $_{mean}P - P_i$ [My] |
| | 9.34737 | 0.56000 | 84.52 | P | theoretical: | difference: | | |
| 0 | 4.73011 | 0.14000 | 81.18 | P/2 | 4.67368 | -0.05643 | -1.21 | -2.57071 |
| 1 | 3.05295 | 0.06000 | 77.71 | P/3 | 3.11579 | 0.06284 | 2.00 | -0.89355 |
| 2 | 2.32542 | 0.03500 | 85.36 | P/4 | 2.33684 | 0.01142 | 0.49 | -0.16602 |
| 3 | 2.02247 | 0.02600 | 80.97 | P/5 | 1.86947 | -0.15300 | -8.19 | 0.13693 |
| 4 | 1.62996 | 0.01700 | 78.41 | P/6 | 1.55789 | -0.07207 | -4.63 | 0.52944 |
| 5 | 1.40048 | 0.01300 | 87.72 | P/7 | 1.33534 | -0.06514 | -4.88 | 0.75892 |
| 6 | 1.07108 | 0.00740 | 74.19 | P/8 | 1.16842 | 0.09734 | 8.33 | 1.08832 |
| 7 | 1.04274 | 0.00700 | 81.58 | P/9 | 1.03860 | -0.00414 | -0.40 | 1.11666 |
| $_{mean}P$ = 2.15940, $\sigma_P$ = 1.23872, 2t-test $P_i$ to $_{theor}P_i$: 0.9716, F-test $P_i$ to $_{theor}P_i$: 0.9891. Practically same population. No outliers (1σ) | | | | | | | | |

The above probe greatly justifies the superseding of the GPTS92, as subsequently done with the CKGPTS95 by Cande and Kent (1995). While the current timescale, as tuned so well to the Earth's precession cycles, has also led many to confuse adynamical (relatively low-energy-) reflections of astronomical periods for dynamical (relatively high-energy-) periodicity, it nevertheless enables pinpointing the culprit, Fig. 2 and Table 2. Note here that, in addition to neglecting to account for reflections as such, one of the main reasons for reflection periods passing for physical cycles lies in an inherent inability of the Fourier and many other methods of spectral analysis to estimate the statistical significance of spectral peaks correctly; see, e.g., Erlykin et al. (2017, 2018). Modern geological timescales that are tuned astronomically — to orbital frequencies primarily — have, with their systematically erroneous nature, hindered paleodata-based quantitative studies of the already obscured role of those same phenomena (precessions primarily) in the origination of global geophysical phenomena like the geomagnetic reversals and Milankovitch theory of astronomical forcing of Earth climate. For example, many who suppress (in the above-described or some other way) the known systematic constituents in the signal, such as the precession, p, do so thinking it would be beneficial in absolute terms. They forget that by doing so, they also destroy or damage any harmonic signal associated with the suppressed constituent. However, such interconnected co-signals are far from useless, like when searching for any harmonic response of a system to external periodic forcing.

## 5. Origin of 2π phase-shift

As shown above, the P-modulation of p is a phase shift by the annual base or full circle, 2π. This unity-timer phasing tuned to annual repetitions of some parameter of the Earth-Moon-Sun orbital system indicates that the origin of the shift as recorded by paleodata is in annual variations of some of the time systems already inherently inconsistent due to differences from interactions between Earth rotation and revolution.

Orbitally forced climate oscillations get recorded in sedimentary archives through changes in sediment properties, fossil communities, chemical, and isotopic characteristics. (Gradstein et al., 2004) Therefore, the impressing mechanism here is simple: the imprinting itself is always and only done via annual cycles, so its timing is subject to any time-system variations from divergences injected into time systems annually. Any yearly variations in the time systems should resemble the Equation of Time — itself reflective of annually varying differences between local mean time and apparent solar time; see, e.g., Seidelmann (1992). The Equation of Time depicts an effect that arises due to the inclination between the planes of the ecliptic and equator (i.e., the obliquity cycle) and the eccentricity of Earth's orbit, i.e., non-uniformity in the apparent motion of the Sun around the ecliptic (i.e., the eccentricity cycle).

The Equation of Time varies through a year in a smoothly periodic manner by up to 16 minutes, Fig. 3, cumulatively introducing secular variations of the Equation of Time into the celestial clockwork. Modern geological timescales are tuned to precessions primarily, and obliquity and planetary precessions as secondary gauges. However, the Equation of Time is virtually independent of precessions and thus mainly unaccounted for by timescales. The minimal discrepancies in timing project annually-secularly in a smoothly changing fashion onto incompatible timescales, year after year, century to century, and *ad infinitum*. When observed as an annual-only phenomenon, i.e., without duration extending to infinity, those changes appear as Earthbound (seasonal). However, observed over geological time scales, their true nature as the desynchronization origin becomes clear from Fig. 3. This poorly understood phenomenon is in various versions described in the literature under many names (as an indication of confusion it created), such as precession index, precession-eccentricity syndrome, and many others; see, e.g., Hinnov (2013) for a long list. However, as shown here, the issue is due to the annual variation in the Equation of Time, Fig. 3, affecting the paleodata sampling process, rather than being due to seasons which is an oversimplification.

That the reflections of the resulting resonant response of the Earth to astronomical forcing indeed tend freely to infinity as expected in an unfixed oscillating system like the Earth-Moon-Sun, can be seen from the above observation of translation symmetry in paleodata in the ~30–~1600 My band (Prokoph & Puetz, 2015). Observed were (Faraday instability ripples in the form of) tripling, doubling, and halving of reflections, their reflections, and reflections of the period reflections. Further rippling is likely undetectable since buried in long-periodic geophysical background noise. As expected again for the above-stated reason, this echo practically extends over the entire geological record, i.e., since the Earth acquired the Moon, ~4450 My ago. Thus, what I found above in the 1–40 My band (from arguably the best available data, reaching ~80 My before the present) should be observable in other 40-My-wide bands, pending quality data from earlier epochs. While Puetz et al. (2014) gave examples of period modulations in paleodata across a variety of scales, including ky small scales, Omerbashich (2022; 2020b) reported a Moon-driven, >M6.3-seismogenic resonance on the hr–day scales as well and showed that a resonance response of Earth to external periodic forcing is detectable and mappable as actual waves in solid matter passing through continuous-GPS (cGPS) stations.

As a paleosample returns with the Earth once a year to the original position as when created, its physical state (more precisely: its relation to physical parameters in the local environment as traced by sidereal time) does not. Instead, the sample's physical state changes slightly each year due to astronomical cycles other than precession. It only returns to its original state (from the epoch when first created) after one complete precessional cycle of 26 ky. However, by then, this variation impressed a cyclicity into the samples and the phasing appearance of its accompanying modulations (mathematical repetitions). As a result, data of mutually diverse types, including those of cratering, deep cores, fossils, geomagnetic field(s), and sediments, all from various epochs, return the same astronomical reflections regardless. However, those reflection periods are not at the same time strictly reflective of the timings of sample creation anymore, but of changing conditions under which the samples had been relating to their year-to-year successive physical states with respect to original local environments as traced by sidereal and apparent times. The longer the data span, the more enduring the impression, so in Earth polarity records, for example, clusters of P-reflections are seen only in the long-period band, as they tend towards the lower frequencies, Fig. 4 (bar line).

Unlike the above-described mechanism for impressing reflections (mainly trailing, i.e., harmonics with higher-order indices), the action by the most energized part of a (by now mechanical) resonance — the driver and its first few harmonics (with lower-order indices) — requires a carrier wave. Carrier waves can deliver cumulative effects of the resonance's destructive angular deflections, crashing into obstacles such as variations in topography and crustal thickness radially and alongside all the propagation paths of the resonance as a process.

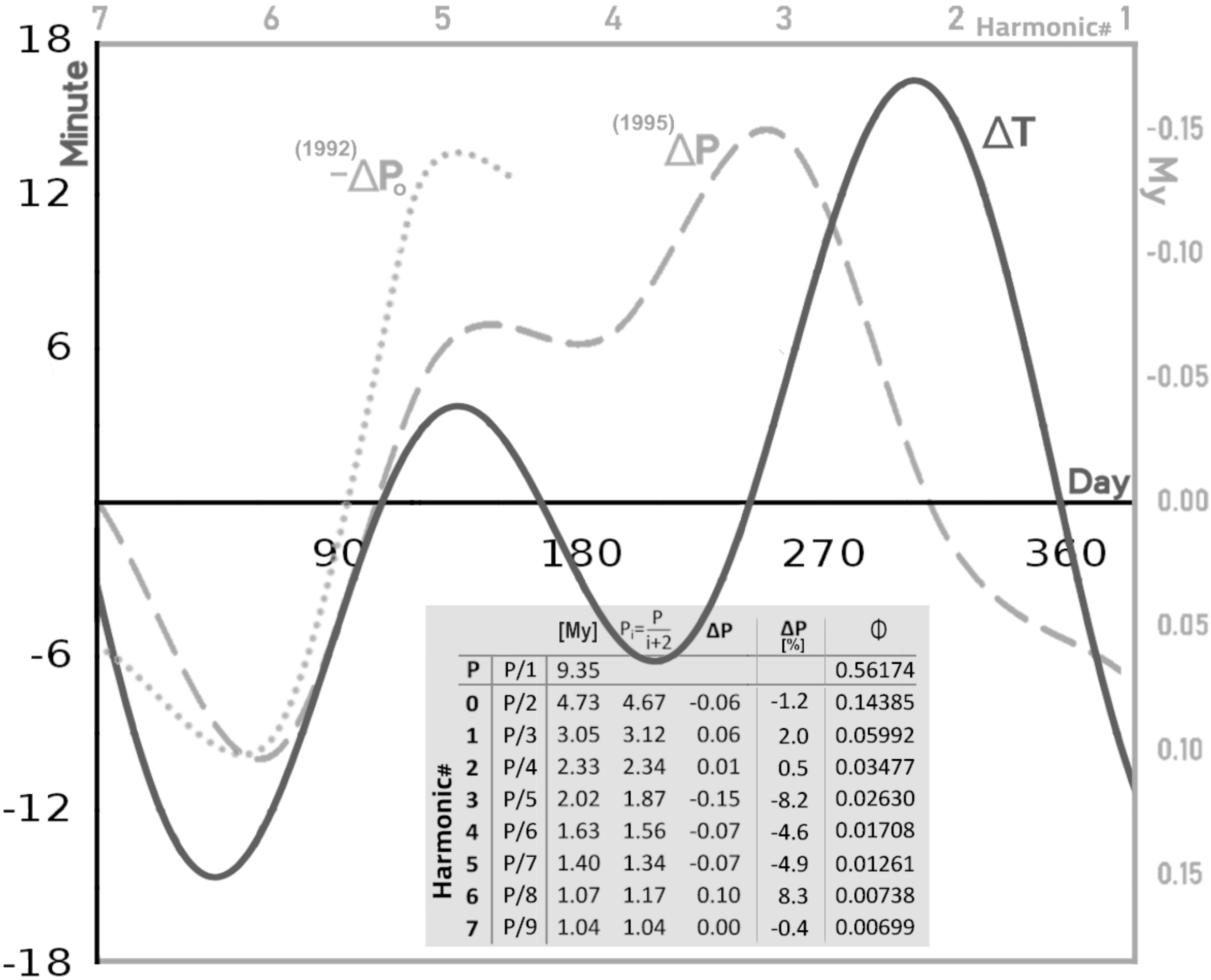

| Harmonic# | | [My] | $P_i=\frac{P}{i+2}$ | ΔP | ΔP [%] | Φ |
|---|---|---|---|---|---|---|
| P | P/1 | 9.35 | | | | 0.56174 |
| 0 | P/2 | 4.73 | 4.67 | -0.06 | -1.2 | 0.14385 |
| 1 | P/3 | 3.05 | 3.12 | 0.06 | 2.0 | 0.05992 |
| 2 | P/4 | 2.33 | 2.34 | 0.01 | 0.5 | 0.03477 |
| 3 | P/5 | 2.02 | 1.87 | -0.15 | -8.2 | 0.02630 |
| 4 | P/6 | 1.63 | 1.56 | -0.07 | -4.6 | 0.01708 |
| 5 | P/7 | 1.40 | 1.34 | -0.07 | -4.9 | 0.01261 |
| 6 | P/8 | 1.07 | 1.17 | 0.10 | 8.3 | 0.00738 |
| 7 | P/9 | 1.04 | 1.04 | 0.00 | -0.4 | 0.00699 |

**Figure 3**. Cumulative projector effect of initial annual timing discrepancies (Equation of Time variation, ΔT, Seidelmann, 1992), solid line, over galactic timescales (long-periodic change, ΔP), dashed line. Here ΔP measures the cumulative change in the ability of paleodata to maintain a theoretical harmonic periodicity with apparent time. The plotted ΔT–ΔP consistency (but also the consistency $\Delta T_i / \Delta T_i+1 - \Delta P_i / \Delta P_i+1$, not shown), along with statistics of the matching, Table 2, revealed that the extracted periods belong to a single process, i.e., represent an ensemble of resonance waves. Annual variations in the Equation of Time are due to imperfections in apparent

times and to astronomical phenomena other than precessions, primarily the obliquity (which then is a candidate for a resonance trigger). Projector errors largely depend upon the initial conditions under which the entrainment has originated and then locked, so paleodata periods' errors reflect those flaws. Inherent CKGPTS95 errors are also due to phase relations between astronomically forced and sedimentary captured timing always being approximate and known accurately only for the last 5–10 My (Hilgen, 1991). Note here that the accumulation occurs alongside neither of the axes, i.e., neither in period nor magnitude. Instead, the accumulating happens across the time domain so that the averaging of the variation sampled by paleodata for a noticeable spectral change to arise happens already on millennial scales. Therefore, it does not take millions of years for the above-depicted projector effect to accumulate to a shape resembling that of ΔT, as this occurs already on millennial scales. That is why higher-order harmonics are approaching ΔT better than lower-order harmonics can. Likewise, the ΔP and ΔT curves eventually coincide for the infinite data span. The superseded CKGPTS92 timescale, $\Delta P_o$ (dotted line), has revealed only the initial ~40% of the projector effect and thus performed relatively worse. Besides, enforcing precessional cycles has left no ≥99%-significant peaks in the GVSA spectrum of the superseded CKGPTS92.

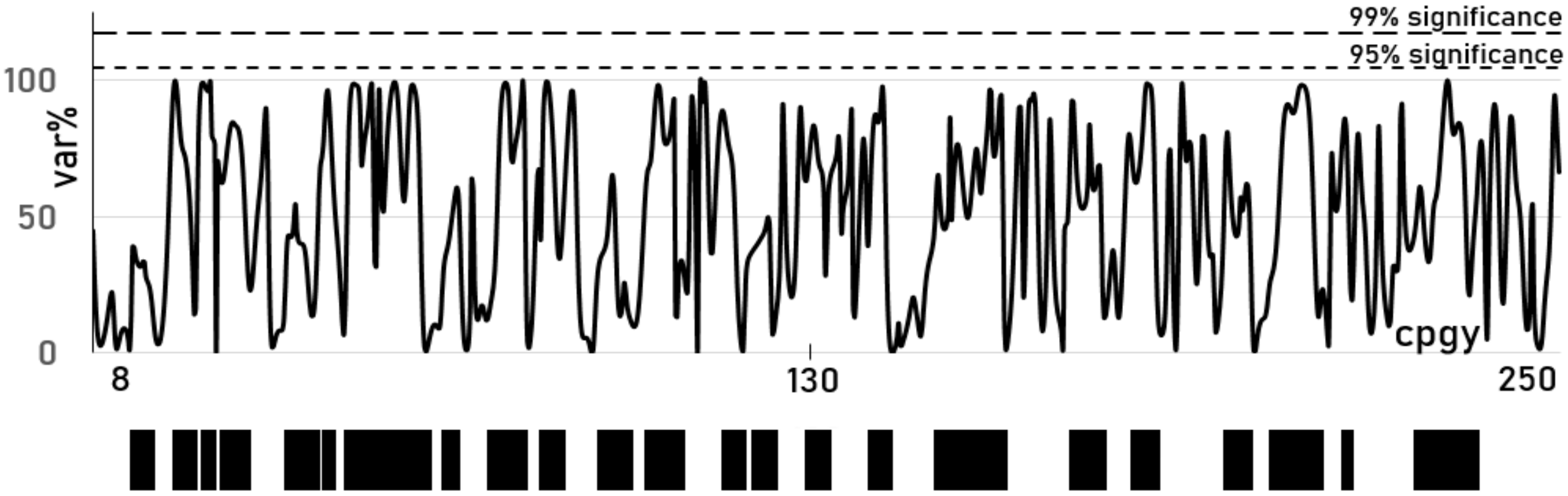

**Figure 4**. The GVSA spectrum of CKGPTS95 timescale calibration, Table 1, after enforcing the modulated driver P, Fig. 2, and the apsidal-precessional period, 29.81 ky, Fig. 1. All ≥99%- and ≥95%-significant periods from Figures 1 & 2 vanish below significance. The axial (strongest) precessional period, p, was not enforced, so the spectrum depicts only direct (p-driven) dynamics due to precession affecting the reversals but not due to precessional resonance anymore. As seen from this separation (of a resonance driver from its resonance), since the remaining dynamics are insignificant, the Earth axial precession alone, with period p, is not responsible for the reversals (sampled by the CKGPTS95 calibration points as the most variable/noticeable segment of the geomagnetic field and thus that timescale's most reliable aspect). However, a uniform background-noise level indicates that another systematic physical process contributes to the dynamics of the reversals. Bar line: the width of the bars corresponds to the local clustering of two or more spectral peaks that did not resolve better than half the spectrum's variance percentage range. The area enveloping white space under a cluster is proportionate to the bar-line blackness and vice-versa. The bar line depicts the change in clustering with a decrease in period length, indicating that a physical process still underlies geomagnetic reversals, albeit weakly. As seen, systematic noise clusters in the band's low end, thus dominating the information. Towards the high end, the systematic noise overpowers all other contents while resolving better (seen as spectral peaks splitting more regularly) due to the absence of other systematic contents.

Any physical system driven by a periodic external field shows a discrete time translation symmetry (Guo et al., 2013). Then rigid subharmonic entrainment can be observed in many-body subharmonic responses (Yao et al., 2020), providing a conceptual framework to interpret Earth-Moon-Sun resonances observed in geophysical data. Indeed, such a response is also characteristic of Faraday instabilities (waves or patterns) (*ibid*.), identified by Omerbashich (2020a) as the Solar system's common polygonal (hexagonal mostly) cratering and patterning. Here entrainment means the mode-locking between coupled oscillators of different periods in which they assume a common period (Strogatz, 2000). The Earth's subharmonic response $P_i$ to the Earth-Moon-Sun system's orbital dynamics, Figs 1 & 2 and Table 2, is one such entrainment seen from the system falling into lockstep with ¼p', or rather ¼ of the obliquity's annually-secularly modulated period, P'; Fig. 5. This find confirms the obliquity as (a major) one of the triggers of Earth's resonant response to external (primary tidal) forcing. Other triggers include undulating topography, varying crustal thickness, uneven distribution of inner masses, and mantle flows. Previously, such a subharmonic response — with the fundamental frequency's ¼-lockstep to the driver frequency — was believed to be characteristic of a discrete (quantum-scale) time crystal only, e.g., Yao & Nayak (2018). However, as shown in the present study, (macroscopic) Faraday-instable multi-rigid-body systems such as our Solar system exhibit this lockstep too, and its extracted value turned out to be ¼ the driver. While Yao et al. (2020) conjectured that the classical time crystals could exist in nature on macroscopic scales, e.g., in Langevin dynamics of molecular systems, Puetz et al. (2014) statistically found that the astronomical and geological cycles could be phase-locked synchronously, with biological cyclicities lagging, but offered no terrestrial cause.

Additionally, discrete time translation symmetry has been experimentally observed in a resonator forced externally by another quarter-wavelength resonator — both as period multiplication but mainly as doubling, halving, and tripling (all in parametrically driven tunable superconducting resonators) and as 2π phase shifts accompanied by intensely magnified radiation (in resonators with frequencies n·ω close to multiples n=2, 3, 4 and 5 of the resonator's fundamental mode) (Svensson et al., 2018). Besides, a global resonance on Earth occurs when the phase delay owed to propagation is proportional to 2π (Nickolaenko & Hayakawa, 2002). Here, the period multiplication (including doubling, halving, and tripling), the 2π phase shifts (of both p and p'), and the ¼ lockstep (to p'), all characterize the present study's find from the GVSA of CKGPTS95 calibration, as well. Detection of all three fundamental properties of a discrete time crystal in macroscopic Faraday instability like the Earth-Moon-Sun system reveals a cross-scale nature of time crystal. The translation symmetry, most often in the form of tripling, doubling, and halving of periods, was previously observed in paleodata in the ~30–~1600 My band; see, e.g., Prokoph and Puetz (2015), while ~3P is often reported in the geophysical literature for 1–40 My band too, most recently by Rampino et al. (2020) as 27.5 My, i.e., 3P to within 2%. However, like with most such reports, that reported period was extracted using inapt techniques (Omerbashich, 2021) that cannot decouple resonances from drivers, resulting in a claim of physical significance and causality by way of galactic motions, among other proposed but unsubstantiated explanations.

The computations thus far showed that the CKGPTS95 timescale, constructed based on the nine adjusted calibration values in Table 1, is highly accurate. By inclusion, the GVSA of the CKGPTS95 has shown that most inaccuracies in the fundamental radiometric decay constant can be ruled out (Cande & Kent, 1992), except for a systematic error in the decay constants used in K/Ar dating (Hilgen, 1991).

## 6. Triggering mechanisms

To identify or rule out any triggering mechanisms of the precessional resonance due to orbital forcing (more precisely: the triggers of Earth mechanical resonance due to energy transfer from orbital periods that include direct and mixed effects of orbital resonances), any individual orbital forcing first must be separated. As mentioned earlier, an inherent feature of variance spectra that GVSA uses is a linear depiction of background noise levels. This depiction streamlines relative spectral computations and analyses. So in the following, I first separate drivers from the resonances. Then I compute spectra of spectra to separate any overlying and underlying dynamics from the respective system (the entire information contents) that presumably is systematically dynamical already.

Enforcing the phase-shifted precessional driver P and the apsidal-precessional cycle p' was sufficient to make all periodicity in the GVSA spectrum of CKGPTS95 vanish below significance, Fig. 4. This vanishing has demonstrated that, according to CKGPTS95, all dominant (significant) cycles over the past 83 My related to polarity reversals were also related to the precessional resonance but not to the main (axial) precession itself. No secondary field dynamics existed that precessional resonance dampened. Moreover, the successful separation of the resonance driver from its resonance, Fig. 4, showed that the slightest disturbances in the precession could also be resonance generators. Finally, the apsidal precession $p_1$ could play a role in stimulating the creation of that disturbance and co-triggering the precessional resonance.

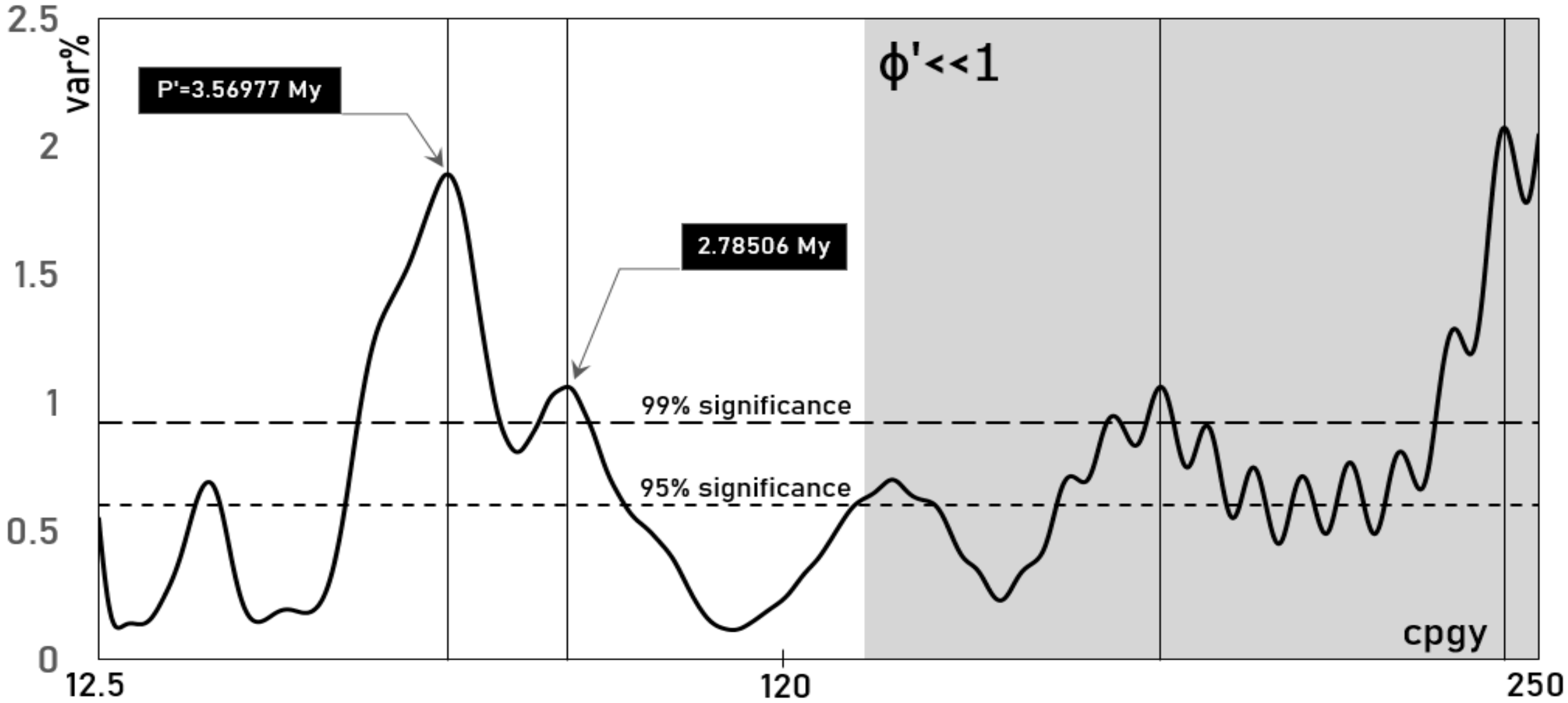

**Figure 5**. The 1–20-My-band GVSA spectrum of the CKGPTS95's precession-free spectrum from Fig. 4. As indicated by their statistical fidelity, only the first two ≥99%-significant periods are of interest. While insufficiently statistically significant to render the two spectral peaks significant physically as well, the relative variance- (thereinafter: power-) dominance in the P'=3.56977 My period makes it alone the Earth's superperiod (superimposed period) that controls other periods in a system. Like with the P relative to p, P' also is 2π-phase-shifted, relative to ¼p', which means that the Earth-Moon-Sun system fell into the ¼ lockstep with the Earth's obliquity that can be viewed then as a co-trigger of the Pi resonance. Unlike a spectrum of a time series, which tells about repetitiveness in data and can thus expose causal physical mechanisms, a spectrum of a spectrum can tell about any repetitiveness of cycles themselves and thus expose overlying system dynamics. Such a procedure in global dynamics studies can have a physical meaning only if dominant dynamics previously are enforced (or ignored, as a more appropriate term given the context). The band was narrowed further to 1–20-My, as methodologically the strictest approach possible, to shift the remaining power to the sub-band of interest; this is justified because floating astrochronologies dominate the data beyond the beginning of Oligocene (~34 My) (Tanner & Lucas, 2014).

This result — that there is no continuous forcing of the reversals, but there are harmonics instead — is supported by Lutz (1985), who found no significant periodicity in geomagnetic reversals and pointed out that long periods in paleodata are either harmonics or not well-defined. Since astronomical tuning left some room in our case for deciding if the precession causes reversals itself, the uniform (i.e., with the highest spectral peaks virtually leveled) noise-background level in Fig. 4 reveals the existence of some other systematic modulation that partakes in the dynamics of reversals. Observing such a modulation mechanism would likely represent the first proof of precessional resonance; see Hinnov (2013) for a review of (orbital) precession resonance studies. Note that orbital resonance among heavenly bodies can indeed cross-scale-translate via energy transfer into mass (particle) resonance on the resonantly affected bodies of mass (Omerbashich, 2020b, 2020a).

Based on statistical analyses, Carbone et al. (2006) claimed that a physical process underlines the geomagnetic polarity reversal phenomenon. If real, such a process could also affect the timescale that here was freed of potentially detrimental effects of astronomical tuning. To test their claim, I use a bar-line descriptor of clustering and plot the change in local clustering with the decrease in period length, Fig. 4. As seen from the clustering getting somewhat resolved towards the spectral band's high end, there does appear to be a physical process underlying geomagnetic reversals. Besides, as expected from systematic noise, the spectrum resolves better (with the decrease in period length), i.e., spectral peaks split more regularly towards the high-end, to a discernable local declustering that then permeates throughout the highest frequencies as their general tendency towards declustering. Therefore, in addition to the virtual flatness of the highest noise peaks, we can see that systematic noise is not lonely in defining spectral information contents. In short, the clustering indicates that the noise appears conjoined by a systematic physical process, thereby corroborating (*ibid*.).

To verify this statistical conclusion further, and in the physics realm primarily, I compute the spectrum of the spectrum from Fig. 4. Spectra of spectra have been used for a long time (for example, to investigate alternate states of matter, namely in demonstrations of the Bose–Einstein condensate). Computing spectrum of a spectrum (rather than comparing individual spectral peaks as done classically) is an inherent ability of GVSA thanks not only to the method's linear representation of spectral background (magnitudes) but also to straightforward statistical analysis incorporated into the spectral computation itself. In this way, and unlike statistical approaches alone (such as reanalyzing spectra of post-enforcing residuals), all modulations — including those arising from combined effects of two or more periods — can be observed or enforced (ignored) simultaneously. Here the 1–40-My-band spectrum of the same-band spectrum of (now precessions-free) CKGPTS95 can tell us about the dynamics of the process overlying the geomagnetic reversals, should such a process exist and be systematic (periodic) in its nature. The spectra computed in the present study apply to any future design modifications of the CKGPTS95 and other geological timescales.

In the 1–20-My band, GVSA of the CKGPTS95 (precession-free) spectrum from Fig. 4 returned a frequency spectrum with six peaks significant at the ≥99% level and eight at ≥95%, see Fig. 5. The lowest spectral peaks at ≥99% were P'=3.56977 My with $\phi$=0.16, and 2.78506 My with $\phi$<0.1. Note here a two-magnitudes-of-order overall drop in spectral magnitudes compared to the CKGPTS95 spectrum itself, Fig. 5 vs. Figs 1–2 & 4, which makes the $\phi$=0.16 of P' indicative of a physical process as well, albeit again an adynamical (or relatively low-energy) one. The 3.5-My period was reported first by Olsen and Kent (1999), who also found its ½ harmonic 1.75-My. They identified those cycles in the large-lake sedimentary record and claimed those periods as the first geological evidence of the chaotic behavior of the inner planets. For them, the 1.75-My (3.5-My) does not correspond to anything in the modulation of precession — but could be a long-period modulation cycle of obliquity, in which case it must come from indirect forcing (*ibid*.). Indeed, as seen in Fig. 5, the spectrum of the (precessions-free) spectral peaks of CKGPTS95, where timescale periods formed a new time series essentially freed from conventional astronomical ties, has revealed P' as a trigger (indirect forcer) of the geomagnetic reversals record. Note that, for a long time, some authors have also offered the obliquity and precession as (semantically) underlying, i.e., controlling variables in the Milankovitch theory as well; see, e.g., Hays et al. (1976).

That P' is another astronomical modulation (now of the obliquity), is seen as before in the case of P (Den Hartog, 1985), so for the 1-yr base oscillation, one obtains theoretically: $\frac{1}{4}\cdot{}^{GVSA}p'_{theor}$=3.56977 My / 360°=9.916 ky, which matches the Earth obliquity's cycle that, as seen here, maintained the lockstep: $\frac{1}{4}p'_{theor}$=41 ky / 4=10.250 ky, to within 3.3%. Since P' was obtained indirectly (from spectra of spectra, to free the data trends of the built-in precessions chronology), the normalized achieved accuracy could be up to an order of magnitude better, to around 3‰. This possibility also follows from looking at the period it overpowered, of 2.78506-My, Fig. 5, which itself is 2π-phase-shifted, i.e., a circular modulation of the 3/10 harmonic of theoretical precession $p_{theor}$=25.772 ky, i.e., 360°·(3/10)p, to within just 0.6‰. Excellent matching to the theoretical precession rather than precession realized in the CKGPTS95-adjusted strata is owing to the above-described methodology, namely combined separation of driver from resonance via computation of the spectra of spectra. Additionally, this overpowering of the precession's second-longest and second-strongest phase-shifted modulation by the obliquity's longest and thus strongest modulation is direct evidence that, in their interplay, it is the obliquity which at least partially controls the precession (and thus indirectly the precessional resonance too) and not the other way around.

Indeed, as seen from Figs 2 & 5, P modulates $P_i$ as a forcer (driver) significantly — so that the $P_i$ are its well-defined harmonics (continuous unity-fractions) rather than reflections (intermittent integer multiples). Then as seen from the spectrum of CKGPTS95 spectrum, Fig. 5, and by the very existence and placement of P', this P' itself is the period superimposed ("superperiod") to P and any P-harmonics. This interplay, in turn, means that the obliquity acts on the precession indirectly: due to the projector effect and the resulting phase shift, Fig. 3, P' impresses upon P instead of p' impressing upon either p or $p_1$ (or both). Note here that the detection of a precessional resonance in the 1–40 My band was enabled in part by methodological band-pass filtering of orbital tuning, which modeled noise in the immediate spectral vicinities of p and ¼p' by focusing spectral power from surrounding frequencies to the Milankovitch bands; see, e.g., Hinnov (2000).

In the same sense, note that the Universal Wave Series (UWS) harmonic periods in the 1-ky–1-My band (Puetz et al., 2016) are a hint at the precessional resonance as demonstrated here to be originating in the 1–40-My band and independent proof that resonance harmonics generally are recoverable from astronomically tuned timescales. Besides, given the methodology used and the number of calibration points, the P' fidelity normalized to at least an order-of-magnitude-larger value does appear sufficient for P' to be called outright physically significant (dynamical) and for its ¼p' base to be recognized as the cause of the chaotic behavior of the inner planets, as inferred by Olsen and Kent (1999). Also, P' is among the most powered periods detected in the present study, right after the axial-precessional period, p, and its circular modulation, P. Then, after using up all the power in data, P stands out as the lead period to some of the most potent terrestrial cycles. Together with its lower-order resonance harmonics, crosscycles primarily (as due to cross-coupled oscillations; the P-harmonics that at the same time are harmonics of other astronomical cycles and their modulations as well), forcing and transformative potential of the precessional resonance-dynamics on inner planets becomes difficult to question.

## 7. Milankovitch theory as a special case of Earth-Moon-Sun resonance dynamics

The Milankovitch theory is successful mostly over the Pleistocene, i.e., between 0.01–2.6 My ago (Feng & Bailer-Jones, 2015). From the resonance dynamics perspective, the reason for this is in the Pleistocene fitting the frequencies of tailing long-period P-harmonics, Table 2, causing the most recent precession's 2π (circular) modulation to be confused for the precession's base cycle and the primary (direct) energy-transfer mechanism. In the same sense, however, P and its first few harmonics carry most of the power of Earth-Moon-Sun resonant dynamics, which is why they affect geophysical processes the most. After all, it is P, its first few harmonics, and their various modulations, which get reported from numerous and seemingly unrelated types of paleodata; see, e.g., Rampino and Prokoph (2020) for an overview, and Omerbashich (2021) for a summary of such adynamical periodicity.

As mentioned earlier, in addition to Carbone et al. (2006), who also saw an underlying process in geomagnetic polarity reversals, Olsen and Kent (1999) also previously reported the obliquity periodicity in the geological record. Those two studies then corroborate the here presented finding of a physical process causing the reversals and excursions (incomplete or short-term reversals). Note again that timescales, in practice, are tied to the precessions primarily (and to a lesser degree eccentricity) because the geological record and other global data do not seem to reflect precession cycles. Similarly, the eccentricity cycle generally is considered insignificant relative to other orbital drivers.

That the results of the present study and the supporting two studies mentioned above are correct is evident from certain aspects of the Milankovitch theory, specifically its predictive abilities. For example, the lack of precession in the geological record historically is attributed to the importance of annual insolation (controlled by the obliquity), with more influence on polar temperatures than seasonal insolation (modulated by the precession) (Naish et al., 2009). Therefore, the statistical and physical significances of any obliquity modulations are not immediately discernable either, mainly due to the methodology of designing a timescale that has to be undesigned then — here by computing spectra of spectra (note that there are numerous other methods too).

Related to that was the above find that the obliquity, non-critically (seen from affecting systematic noise content right below precessional) and indirectly (as overlying other astronomical cycles), partakes in the control of geomagnetic reversals. Then this finding also supports the Milankovitch (untestable; see *ibid*.) hypothesis that summer half-year insolation intensity, with its sensitivity to the precession, can control geophysical parameters like the surface temperature and growth and decay of ice sheets. Therefore, the 3.5-My cycle confirmed in the present study is a likely candidate for the Earth's superperiod as a periodicity superimposed on all other periods in a physical system. To show in the latter case that a superperiod has a physical meaning, any adynamical data periodicity or reflections must first be separated from respective drivers. Only after such signal separation, here used to simultaneously remove any effects of the astrochronology from the timescale calibration, can it be expected from the spectrum of the spectrum of that timescale to extract a meaningful (dynamics-related) repetitiveness of a data period itself.

Also, the present study supports the Milankovitch theory in that it did not find the 100-ky eccentricity period like that in time-series of ice ages — proclaimed by some as the period which the Milankovitch theory cannot explain and therefore collapses as a theory; see, e.g., Berger (2012). Moreover, there is no phase-shifted eccentricity either. However, a faint although ≥99%-significant spectral peak at 100.02-ky with a very high 99.5-var% magnitude but forbiddingly negligible Φ=0.000064 is present, but only when the upper limit of the spectral band of interest is set very near the precession and obliquity periods, say in-between those at 0.035 My so to cancel out any spectral leakage effects. (For example, setting the spectral band's upper limit just below obliquity as another dominant cycle while leaving the opposite vicinity vacant demonstrates an unmitigated leakage effect: the only ≥99%-significant peak at 84.02-ky (twice the obliquity to within <2.5%) gets picked up at a very high 93.2-var% magnitude, but again with a forbidding Φ=0.000045. Thus the obliquity cycle itself leaks beyond detection, and all that remains is its 2T reflection.)

This proven presence of a period only after a deliberate data manipulation identifies the 100.02-ky as an obliquity-precessional resonance period (also note its suspiciously round precision, indicating an artificial, i.e., precession-tuning origin rather than a genuine eccentricity cycle), and shows that its detection by other workers is likely due to data preprocessing including padding, filtering, and tapering, especially when using Fourier methods. This separation of an astronomical period from a similar but unrelated resonant period was possible thanks to the earlier described and repeatedly demonstrated absolute-extraction abilities of GVSA from as few as three values.

Insomuch as ice ages and geomagnetism are phenomena not entirely independent, this all-embracing indication of positive verification of the Milankovitch theory is owing to the most profitable use of data by definition of a scientific endeavor: in their raw form. Berger (2012), who even used a GVSA approximative variant to compute a short-band periodogram of the geomagnetic record compiled by Bassinot et al. (1994), supports this conclusion. Berger (2012) has thus found no 100-ky period either, while also strongly suggesting significant variation near obliquity and in various lines related to precession (*ibid.*), resembling resonance interplay. This outcome was to be expected if the Milankovitch forcing — and the precession especially — is the Earth's leading dynamical driver, say, of ice mass, see (*ibid.*).

## 8. Verification of resonance-moderation of polarity reversals and strata

To verify the above detection of the mechanical-resonant response of the Earth to external forcing due to entrainment, along with the consequent resonance-moderation of magnetic polarity reversals and strata, I first demonstrate the ability of GVSA to extract known resonances from paleodata. Subsequently, I show that the Rampino period, $P_R$, while being the sole and dynamical actor in geomagnetic reversals, belongs to the precessional resonance as well; $P_R$ is a carrier wave of resonance's destructive deflections, indirectly responsible for the polarity reversals by downward penetrating, enveloping the inner core, and eventually flipping the geomagnetic polarity.

**8–a. Extraction of previously reported planetary resonances**

For that purpose, I compute the 1–40-My-band frequency spectrum of the time series of residuals remaining after enforcing the precessions in the CKGPTS95 calibration. Residuals analysis is another feature inherent to GVSA that makes this method a comprehensive numerical-statistical analysis package rather than just a numerical computations algorithm. Here, spectra of a residual time series can unhide any periods previously suppressed by a systematic signal being enforced (Wells et al., 1985). (Note that here the entire resonance train got enforced with enforcing P; this is another, now methodological, evidence that the ensemble of $P_i$ harmonics as a whole indeed forms the precessional resonance.) Thus the GVSA spectrum of residuals has revealed ≥95%-significant known periods of Earth-Mars planetary resonances recognized earlier in various (mostly older than 83 My) parts of the geological record and now — for the first time — also in the 0–83-My portion (see Table 3 of Hinnov, 2013); see Table 3 for complete data.

**Table 3**. Matching of ≥95%-significant periods in the 1–40-My-band GVSA spectrum of CKGPTS95-spectrum's residuals remaining after enforcing P (and thus $P_i$, too) against the same periods as reported previously; see Table 3 of Hinnov (2013) and Table 1 of Rampino and Prokoph (2020). There were no ≥99%-significant peaks — as expected after enforcing the dominant precessions and their reflections (precessional-resonance) signals in data naturally dominated by resonances (not just the enforcing-suppressed ones, but also planetary like Earth-Mars). Note a very low statistical fidelity, Φ, on most periods, indicative of resonant or otherwise mostly powerless spectral periods. Here power is in the sense of relayed energy, not GVSA magnitudes as commonly given in variance percentages (but which can also be expressed in spectral power (spectral density) units of dB, see Pagiatakis, 1999). The CKGPTS95 timescale spans 0–83 My, Table 1.

| Stratigraphic record | Planetary resonance | | GVSA of CKGPTS95-spectrum's residuals (0–83 My), p & P enforced | | |
|---|---|---|---|---|---|
| | g4-g3<br>[My] | s4-s3<br>[My] | period<br>[My] | $\Phi_{period}$ | magnitude<br>[var%] |
| NEOGENE<br>Pleistocene–Miocene (0–9 Ma)<br>$\delta^{18}O$ sea level<br>Miller et al. (2005); Boulila et al. (2011) | | 1.2 | 1.11905 | 0.00110 | 73.75 |
| MIOCENE-OLIGOCENE (20–34 Ma)<br>Benthic marine $\delta^{18}O$<br>Ocean Drilling Program (ODP) Site 1218<br>Pälike et al. (2006); Boulila et al. (2011) | | 1.2 | 1.11905 | 0.00110 | 73.75 |
| CRETACEOUS<br>Aptian–Albian (100–125 My)<br>Scisti a Fucoidi, Italy<br>Grippo et al. (2004); Huang et al. (2010) | 1.5 | | 1.60695 | 0.00230 | 87.69 |
| TRIASSIC–JURASSIC<br>Norian-Pliensbachian (205–184 My)<br>Inuyama Chert, Japan<br>Ikeda and Tada (2013) | 1.6–1.8 | | 1.60695<br>1.83749 | 0.00230<br>0.00300 | 87.69<br>82.73 |
| TRIASSIC<br>Carnian–Rhaetian (230–200 My)<br>Newark Series, USA<br>Olsen (2010) | 1.7 | | 1.60695<br>1.83749 | 0.00230<br>0.00300 | 87.69<br>82.73 |
| TRIASSIC<br>Anisian–Ladinian (245–237 My)<br>Inuyama Chert, Japan<br>Ikeda et al. (2010) | 1.8 | | 1.83749 | 0.00300 | 82.73 |
| PERMIAN<br>Wuchiapingian–Changhsingian (251–260 My)<br>Wujiaping-Dalong Formations<br>Wu et al. (2013) | | 3.11 | 3.18584 | 0.00910 | 76.29 |
| Compilations | Average lead cycle<br>[My] | | GVSA of CKGPTS95-spectrum's residuals (0–83 My), p & P enforced | | |
| ENDING WITH PALEOPROTEROZOIC<br>Holocene–Orosirian (0–2023 My)<br>Compilation of 58 cratering & 35 mass-extinction reports<br>Rampino and Prokoph (2020) | 26.5 (mass extinctions)<br>~26 (craters + mass ext.) | | 26.53386 | 0.63000 | 77.02 |
| Various paleodata | period<br>[My] | | GVSA of CKGPTS95-spectrum's residuals (0–83 My), p & P enforced | | |
| PERMIAN/TRIASSIC<br>Holocene–Permian/Triassic (0–253 My)<br>Mass-extinction episodes<br>Raup and Sepkoski (1984) | 26.4 | | 26.53386 | 0.63000 | 77.02 |
| ENDING WITH PALEOPROTEROZOIC<br>Holocene–Orosirian (0–2023 My)<br>Cratering record<br>Chang and Moon (2005) | 26.4 | | 26.53386 | 0.63000 | 77.02 |

**8–b. Extraction of periods previously reported from cratering and extinctions**

I now extract periods reported previously as dominant in cratering and mass extinction records. Note that the most significant period revealed from the precession-free geomagnetic reversals CKGPTS95 calibration, of 26.5 My and at a very high magnitude of 77.0 var% with a moderate Φ=0.63 (nonetheless relatively highest of any period extracted in the present study), could not be identified as a resonance reflection or harmonic. This period has been reported previously by Rampino and Prokoph (2020) as the average dominant cycle in the first comprehensive compilation of 58 reports of cratering and 35 reports of mass extinctions and here, therefore, termed the *Rampino period*, $P_R$. (Note that Raup and Sepkoski (1984) and Chang and Moon (2005) came significantly close to the same value from an extinctions record spanning 0–253 My and a cratering record spanning 0–150 My, respectively, Table 3.) Compiled periods were based on numerous entirely dissimilar methods and approaches to computing and analyzing spectra. This period reveals a powerful wave (the sole and dynamical actor) as it affected the solid Earth. Besides, it is of the highest fidelity found on any period extracted in the present study. Its remarkable power is stressed even further when we know it took the entire precessional resonance train (ensemble) to hide it.

$P_R$ is so pure and persistent that it appears even as a co-driver of other (the short-period; ky-) harmonics, Table 4. However, it does not emit longer-period harmonics, revealing that the resonance confines it to (many) relatively short intervals of time — disallowing it by the nature of the resonant process itself to be ever encountered by the entire geological record. Since also supported by the combined cratering-extinctions record and the extinctions record alone, $P_R$ most likely represents the precessional resonance's crosscyclic (multi-harmonic) carrier of the peak angular deflection. Here cratering is mainly understood as petrified evidence of large-scale polygonal geomorphology seen throughout the Solar system due to the imprinting of actual solid-matter resonance waves via macroscale Faraday latticing into the molten material (melted on impact as well, in a snapshot fashion). In the same sense, originally polygonal crater edges eventually erode into circular shapes (here the most realistic scenario) rather than the other way around, i.e., from circular into polygonal geometries as believed by some (the least plausible scenario since erosion takes much longer time than melt impression); see Omerbashich (2020a).

The decisive role of precessional mechanical resonance for Earth is particularly plausible since found in both the cratering and the calibration as the most accurate portion of the currently accepted (timescale of the) geomagnetic polarity reversals record. These multiple detections from seemingly disparate data mean that the precessional resonance is directly, and the precession and obliquity in tandem indirectly (along with geomorphology varying in crustal thickness), responsible also for the geomagnetic reversals. Finally, Rampino and Prokoph (2020) deduced that records of cratering and mass extinctions are somehow connected. Indeed, their fundamental connection follows from the above computational result directly: not only is the precessional resonance long-term destructive (mainly causing geomagnetic excursions — started but never completed reversals — that therefore appear as gradual processes while masking the cataclysmic nature of the reversals) but also strictly terrestrial and incessant. Then the record of mass extinctions is a record of the destruction of evidence of life during *Transformative Resonant Events* (TRE) instead of the periodicity of mass extinctions themselves (note that only evidence of a few remarkable mass-extinction events is beyond doubt in geology). This unavoidable inference questions most, if not all, past claims of periodic mass extinctions.

The well-established fact that life always appeared to have sprung back incredibly fast following an extinction event (*ibid*.) inevitably reveals the circular logic of the proponents of periodic mass extinctions, making the mass-extinction records work just the same in the opposite direction — as evidence of the no-mass-extinction scenario of life on Earth. Besides, the fact that (35 globally random) extinction episodes can average to the value of the dominant period of geomagnetic reversals and that a combined record (of 35 global extinction events and 58 cratering events) does so within half-order of magnitude is remarkable itself because it reveals that all three phenomena — polarity reversals, polygonal "cratering" (commonly misunderstood as always and only the impact cratering), and "periodic" extinctions are due to a single process. (Here, extinctions are presumed regardless if represented by the record of extinctions or their sample's violent decimation like what large igneous provinces had done to the geological record, e.g., Prokoph and Puetz, 2015).

Most importantly, since the average of dominant periods, as returned from globally randomly sampled and diverse data sets, match the dominant periodicity of one of those global phenomena exactly, the underlying physical process must be ergodic. Thus, there is nothing chaotic about those processes, including the geomagnetic field whose recent ergodicity is established (De Santis et al., 2011). Then extraction of the Rampino period means a data-based confirmation of geomagnetism's overall ergodicity. The data of alleged (periodic or otherwise) extinctions support this possibility: the average intensity of extinction of marine life during the Phanerozoic has decreased, i.e., changed on a steady downward trend, while the number of families increased unobstructed; see Figs. 1 & 2 of Newman and Sibani (1999). Namely, Fig. 1 in (*ibid*.) shows the extinction rate in families per My for marine organisms as a function of time (callout: origination rate for the marine organisms), while Fig. 2 in (*ibid*.) depicts the number of known families of Phanerozoic marine organisms, as a function of time (linear trend dashed; exponential trend dashed) (callout: the integrated percentage extinction of families). Note here that trends are steady but non-monotonic and, due to sensitivity to the ocean tidal component as the most persistent systematic noise constituent, marine genera somewhat exaggerate the trending. Thus it is implausible that most, if not all, species had developed over time natural resistance of the same kind and at the same level or extent to the causes of extinction since that would imply thousands of factors to coincide and remain congruent over half a billion years at least. This steady drop in extinction intensity resembles energy dissipation of an underlying physical process, and the increase in the number of families (and lifespan too, as later studies have shown), both correspond to a scenario without any extinctions whatsoever, i.e., as if geophysical upheaval responsible for geomagnetic reversals and cratering has also decimated the record of species but not the species themselves.

The orbital-mechanical resonance, which arises due to energy transfer, is a process that fits the above description: it naturally dissipates over very long (hundreds-of-My-) intervals, only to rebound at other times, all while destroying (via harmonics or $P_R$) any petrified evidence of speciation systematically-periodically. The conclusion that the constant drop is a sign of an underlying resonance is supported by a steady global in-size reduction of polygonal (resonant) geomorphology, such as patterning and cratering. Specifically, such geomorphology has changed from ancient kilometre-scale shapes like the polygonal (mostly hexagonal) craters of Faraday latticing seen throughout the Solar system to metre-scale shapes like the patterns in present-day salt deserts on Earth; see Omerbashich (2020a) and Lasser (2019), respectively.

**Table 4**. Values of the ≥95%-significant GVSA spectral peaks from the 0.02–40-My spectral band of the geomagnetic reversals' currently accepted timescale, CKGPTS95, Table 1. The ≥99%-significant periods are plotted in Fig. 1. Compared to Table 2 (showing the same but in the 1–40-My band), the extraction of the above periods was affected by the power of the precession cycle, p, as one of the clearest periods found in the present study. The four clusters of split periods are shown as declustered (represented by one period per cluster). The above-listed, short (ky-) periods are mostly harmonics of P – circular modulation of precession p; P' – circular modulation of ¼p' obliquity; and $P_R$ – Rampino period, of 26.5 My (Rampino & Prokoph, 2020). Note that a precessional 20.41-ky period and its 5/2-modulation to within <1%, 50.54-ky, are crosscycles of both Rampino harmonics and precessional circular modulation's harmonics (making it clear that P dominates $P_R$ rather than the other way around). This direct P-$P_R$ connection confirms $P_R$ as the most relevant and potent for orbital-mechanical resonance energy transfer. Note that the most efficient feature of mechanical resonance in terms of the ability to relay destruction is its magnification effect via frequency demultiplication, which upsurges the energy resonantly injected into a physical system by 100s of times (Den Hartog, 1985).

| GVSA of CKGPTS95-spectrum's residuals (p and P enforced) | | | harmonic |
|---|---|---|---|
| period [My] | $\Phi_{period}$ | mag [var%] | |
| 0.14974 | 0.0001400 | 78.61 | |
| 0.09775 | 0.0000610 | 75.07 | |
| 0.06977 | 0.0000310 | 78.46 | P/134 |
| 0.06929 | 0.0000310 | 74.59 | P/135 |
| 0.06543 | 0.0000280 | 73.49 | |
| 0.05054 | 0.0000160 | 74.34 | P/185, $P_R$/525 |
| 0.04267 | 0.0000120 | 77.72 | P/219 |
| 0.04010 | 0.0000100 | 73.32 | |
| 0.03863 | 0.0000096 | 85.65 | P/242 |
| 0.03346 | 0.0000072 | 79.57 | $P_R$/793 |
| 0.03334 | 0.0000071 | 86.19 | P'/107 |
| 0.02951 | 0.0000056 | 89.23 | P'/121 |
| 0.02422 | 0.0000038 | 78.45 | P/386 |
| 0.02041 | 0.0000027 | 73.50 | P/458, $P_R$/1300 |
| 0.02037 | 0.0000027 | 75.38 | |

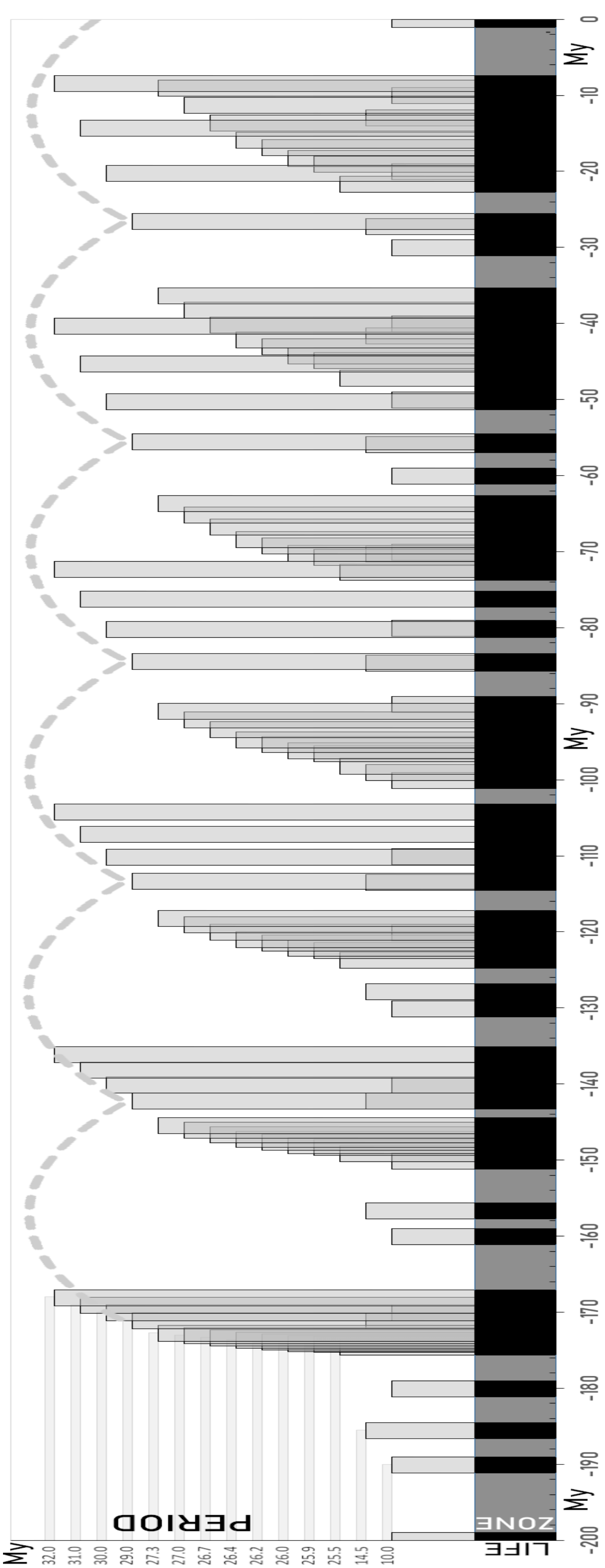

**Figure 6**. Time-domain-superimposing all claims of periods in mass extinctions, as compiled by Rampino and Prokoph (2020) (Table 1 in there), reveals the virtual impossibility of life on Earth. All periods were used (regardless of statistical significance) due to the inherent inability of Fourier and other classical methods for spectral analysis to correctly estimate statistical significance on spectral peaks (Erlykin et al., 2017, 2018). The plot reveals the impossibility of life on Earth in the Phanerozoic (the last half a billion years). The point of origin was selected at -200 My, as being around the middle of the Phanerozoic. The plot assumes all life originated at an instant or during a short interval (of several My). Bar line: Grayness represents life, and blackness is the impossibility of life. Bin spreads fixed at 1 My represent macroevolutionary bursts (minimum time required for a single evolutionary step to develop adaptively in a species) per Uyeda et al. (2011). The 10-My bin is a gauge best fitting the overlap intervals, averaging the time for which most species continue after an extinction event (cf., Barnes et al. (2021) who found that most species continue for more than 30 My past an extinction, while many drops in diversity without recovery are not associated with mass extinction events at all). The dashed arc depicts a resonance period (wave) as it rolls out in the time domain. Based on the varying spread in overlapping, the plot rules out galactic cyclic causes of mass extinctions, revealing that life on Earth would be practically impossible if most of the reports alleging periodic mass extinctions were results of using reliable data and proper preparation and processing techniques. Instead, the precessional resonance via its carriers of destructive angular deflections (here the 26.5-My Rampino period, Table 3, and lower-order harmonics) causes polarity reversals, on Earth during TREs while simultaneously decimating strata — including the record of (thus “periodic”) mass extinctions.

Finally, superimposing in Fig. 6 of all previously alleged mass-extinction periods in the time domain reveals that either life during Phanerozoic (including today) could not exist or the periodic mass-extinctions are not real. Note that the obliquity-induced precessional resonance does not outright discard mass extinctions as such, like the Permian–Triassic Extinction Event. Instead, the record of evolution gets decimated together with, or within a short time from, topography-reformatting TREs. This reasoning agrees with the recent finding that most species continue after an extinction event for more than 30 My, while many post-event drops in diversity without recovery are not associated with mass extinction events (Barnes et al., 2021). In addition, pre-Milankovitch views that climate periodicity is due to changes in Earth's magnetic field (Puetz et al., 2016; Raup, 1985) made sense only due to the geomagnetism record's dominant bias in the same way. Then geomagnetism polarity reversals, polygonal "cratering", and "periodic" mass-extinction records describe the same phenomenon of obliquity-triggered precessional mechanical resonance that causes stratigraphically recorded changes — in geomagnetism and cratering primarily. This recording is due to actual resonance waves in the solid matter (Faraday latticing), but only apparent changes in evolution — as evolutionary evidence (residing embedded within solids) gets exposed to damaging resonance waves. Note here that, while the geomagnetism record gets decimated too, it is the least affected type of solid-matter-residing paleodata since only simultaneous total (rarest) reshaping of entire continents can alter that record significantly in terms of its systematic information contents.

Other examples of GVSA's ability to compute resonances in data at absolute accuracy and ≥99% or ≥95% significance levels include resonance extractions from time-series of $M_w$5.6+ earthquakes' occurrences for solid Earth (Omerbashich, 2020b) and from moonquakes for solid Moon (Omerbashich, 2020a), both cases also reporting ≥89%- and ≥67%-significant harmonics as well, all the way up to 1/72 of the forcer in the case of solid Earth. Besides, Omerbashich (2022) showed from continuous GPS data that GVSA can absolutely (i.e., to the accuracy of the measurements) accurately extract solid-matter resonance as it forms actual particle waves. Worth noting here is the main difference between detecting orbital resonances transferred directly (mechanically and by way of gravity) into the solid matter along particle waves and those impressed into gaseous matter indirectly as a temporally undersampled powerless signal: while in the latter case, characteristic of various types of paleodata, statistical fidelity of so extracted resonance periods is typically $\Phi<12$ and often $\Phi<<1$ and even $\Phi<<<1$, in the former case it is regularly $\Phi\geq12$ and often $\Phi>>12$ and even $\Phi>>>12$. For example, while dealing with >99% populated time-series, (*ibid.*) found in the extractions that fidelity occupied the 12–11 ranges on the shortest and had reached the ranges of values as high as 60,000–9,000 on the longest periods.

## 9. Identification of moderation carrier

Next, I examine the phase relations of $P_R$ to P to verify that the Rampino period, $P_R$, does represent the precessional resonance's crosscyclic (multi-harmonic) carrier of the peak angular deflection. Should such a causal relation exist, it would expose $P_R$ as a modulation of the already phase-shifted precession p. In that case, $P_R$ would also be the primary modulation of P, and thereby the only sole actor besides ($2\pi$p, $P_i$) resonance in the 1–40 My band.

Since not an orbital period itself, we cannot expect $P_R$ also $2\pi$-phase shifted, i.e., in the same way as precession and obliquity did. Then, since $P_R$ was extracted as a sole actor only once — in the absence of (p, $P_i$), how P modulates $P_R$ is determined entirely by the effects of annual variation in the Equation of Time on P itself, Fig. 3. Puetz et al. (2016) already showed an effect this variation has on P, albeit in the 1-ky–1-My band and for the base cycle in that band, as a phase coefficient k=2.829. (Note that those authors, based on observations of the discrete time translation symmetry alone (i.e., of paleodata periods tripling, halving, and doubling), mistakenly assigned the unit of a solar year to the above coefficient of proportionality and called it a period, $P_{0,0}$, when in fact it is the symmetry alone which arises due to annual variation in the Equation of Time, Fig. 3).

Since $P_R$ is tacitly assumed to be itself affected by the same variation and we want to test an implied physical hypothesis (of $P_R$ being the strongest resonance period in terms of damage potency to solid matter, on a par with P itself or even more robust), k is already a phase-shift of $P_R$ and is a (dimensionless) coefficient. Then to establish or rule out the above hypothesis, it suffices to compare the match of $P_R$ and P against k. Furthermore, since we can see that $P_R/P=2.838$, the matching is within 3‰, or practically absolute, thus independently and numerically validating the results from GVSA, Fig. 3.

Furthermore, since k itself (again, as a coefficient instead of period) is just a reflection of π, i.e., k=9/10·π to within 0.5‰, i.e., practically 1·π, this odd-multiple π-phase-shift of $P_R$ from P reveals that $P_R$ is a carrier of classical destructive rather than constructive (even-multiple) interference, itself phase-shifted from P so that $P_R=(9/10)\pi\cdot P$. The instability in the k=π equality due to a 9/10 reflection alone (itself a reflection to within 0.5‰) is an independent confirmation that the destructive interference of $P_R$ is due to inherently unstable angular deflection of mechanical resonance.

Thus $P_R$ originates as the system period, i.e., around the mid-band and as a $(9/10)\pi\cdot P$ modification itself — as the result of an interplay of all resonance harmonics, i.e., the above-noted entire train, Fig. 2 and Table 2. This interplay is why $P_R$, while phase-shifted itself (albeit not by 2π), forces (or relays, rather) its higher-order harmonics, Table 4, revealing its terrestrial rather than galactic or other origins, and P as the resonance driver. Although an excess critical speed and angular deflection make for the potentially destructive side of any mechanical resonance, the most potentially damaging feature of such resonance in the case of closed physical systems like the Earth-Moon-Sun is far more critical. Namely, it brings about the mentioned resonance-magnification effect via frequency demultiplication that can magnify energy injected resonantly into a physical system by 100s of times (Omerbashich, 2007a; Den Hartog, 1985). In a closed physical system like a planet, this magnification gets channeled via a carrier wave such as $P_R$.

Such a sole-actor, energetic period like $P_R$ can co-instigate geomagnetic polarity reversals and be capable of upheavals on and inside the Earth. Such cataclysms include resonant (and therefore mixed, intermittently periodic-quasiperiodic), all-transforming events like plate tectonics or reformatting of continents, which also significantly alter strata. However, the complete record of geomagnetic polarity reversals is not periodic with $P_R$. This apparent absence is due to the varying ability (to complete inability) of incessant resonance train's member waves to move across vast distances and depths. These individual harmonics partake individually during TREs in decimating strata and other records of paleodata, thus adding an appearance of pseudoperiodicity or chaos in data. The variation in an individual harmonic's ability to move as a wave depends upon which resonance frequency turns into the largest destroyer-wave feeder at a given time and location. Then identifying and ignoring (turning on and off) those waves in all combinations can help map the underlying dynamics in space and time with the aid of spectral analyses. The significant alteration of strata by a $P_R$ wave during a TRE, and then the downward continuation of that wave and its harmonics, and then penetrating and finally enveloping the inner core, is the core/polarity reversal mechanism.

Numerous models support the results of the present study in some of its aspects at least. Thus, a model by Pétrélis et al. (2011) supports the possibility that such a mechanism is at play by establishing a link between plate tectonics and reversal frequency, i.e., a correlation between lopsided continental geomorphology and the soon-following reversals. While corroborating the Carbone et al. (2006) finding of reversals' clustering, a model by Stefani et al. (2007) agreed well with the CKGPTS95 and identified resonant features in geomagnetic polarity reversals. On the other hand, while unable to confirm the clustering, a core-rotation model by Hoyng and Duistermaat (2004) has suggested that polarity reversals are relatively fast events. In addition to numerical and analytical modeling, numerous experiments addressed the geomagnetic reversals issue. Hence, dynamos generated by mechanical driving like precession or tidal forcing have become popular in recent years, and experiments focusing on precession-driven flows are conducted in various labs (Giesecke et al., 2019).

The high accuracy and precision of computations of previously reported periodicity, besides owing advantage to the GVSA, also mean that virtually all ~3P periodicity previously reported from paleodata most likely are the underestimated and overestimated extractions of $P_R$, as evading detection is one of the characteristics of resonance carrier waves. Besides, while indeed detecting a ~3P period in paleodata, many of such reports also found no statistical significance in it (Table 1 of Rampino and Prokoph, 2020), and variable significance is another characteristic of resonance carrier waves (in addition to being characteristic of using inapt computational tools and approaches, as described above).

As extracted in the present study, $P_R$ is matched practically by the average from all the estimates of dominant periods from the extinction record. However, the cratering record's average is significantly mismatched (*ibid*.). This difference was as expected since encapsulated or sheltered strata preserve histories of natural disasters more faithfully than the craters exposed to weather and erosion ever could. Previous reports corroborated the finding of $P_R$ in geomagnetic reversals in the present study, e.g., by Raup (1985), who, in a coarse study, found geomagnetic reversals periodic at 30 My, which probably was an overestimated $P_R$.

## 10. Discussion

Modern approaches to dating events and fossils in the geological record over the past half a century have relied heavily on tuning that record to repeatable astronomical events in the Solar system — especially so for missing ends of strata series or otherwise obscured parts such as most of paleodata older than ~40 My. However, as with any modeling, this contrived "clockwork geoscheme" is only as good as end-users' awareness of all its drawbacks and potential error sources. It took only a few decades for ad-hoc tying arbitrarily selected portions and ranges of the record to celestial mechanics to diverge into the realm of spectral analyses of so-tuned datasets declaring practically every adynamical reflection of astronomical cycles as verification of some genuinely natural process. For instance, spectral analyses of diversity in paleodata often invoke even extraterrestrial causal mechanisms, such as galactic or universal, as the cause of mass extinctions — thereby implied as repeating themselves. However, such explanations are not convincing since they do not explain the appearance of equally justifiable and possibly related sub-harmonics in the spectral analysis (Baker & Flood, 2015). Not only do extraterrestrial causes need not be invoked to explain widely present periodicity in paleodata, but those periods themselves, as extracted, are also suspect due to undersampling and overestimating issues pervading analyses of the geological record. For example, as seen in Fig. 6, periods of so-called periodic mass extinctions superimposed in the time domain make most of life on Earth seem virtually impossible.

By focusing on the rarely probed 1–40-My long band, it became possible in the present study to mathematically avoid the dominance of the axial, p, and apsidal, p1, precessions, Fig. 1 — amplified and focused in the process of astronomical tuning of the CKGPTS95 timescale of geomagnetic reversals. This approach has revealed an annually-secularly modulated precession of P=9.34737 My, along with its complete ensemble of harmonics in Fig. 2: $P_i=P/i$, $i=1...9$; $i\in\mathbb{Z}^+$, Fig. 2. The $P=P_1=360°\cdot p$ is the lead (annual) modulation of precession harmonics due to the annual variation in the Equation of Time, Fig. 3. Rather than cumbersomely comparing individual peaks in the timescale spectrum to discern the possible modulation impresser or trigger, computing the spectrum of that timescale calibration's spectrum in the same band instead reveals the obliquity's annual modulation, P'=3.56977 My = $360°\cdot$¼p'. Here, ¼p'=9.916 ky or the obliquity's base cycle ¼$p'_{theor}$=41000/4 to within 3%, overlies the entire record then and triggers the here extracted precessional resonance.

Hence, when spectrally analyzing paleodata, it is not sufficient to declare a statistically significant period also physically significant or outright discard an extracted period that is statistically insignificant. Such detections warrant further investigation into possible mutual mathematical relationships among statistically significant periods. Suppose a detected period is part of a modulation ensemble, i.e., one of (usually many) mutually mathematically directly related periods. In that case, such a batch represents n- reflections (integer multiples of the shortest period — the driver) or harmonics (integer splits of the longest period — the forcer) or unspecified modulations (ratios of the driver or the forcer). Reflections also can appear as forcers of other modulations, which are then termed reflections and not harmonics, as the term harmonic gets reserved for possibly dynamical (relatively high-energy) cyclic physical phenomena. If the driver or forcer turns out to be astronomical — like the Earth's axial precession period of ~26 ky — then its ensemble of modulations is itself physically insignificant or meaningless, too. Only after enforcing of entire such ensemble (where it suffices to ignore just the driver of forcer for its entire ensemble to get ignored too) can one begin a quest for periods as signatures of natural forces or phenomena that had dynamically affected the data being examined.

Then additional criteria are required to determine the physical meaningfulness of statistically significant periods. As mentioned earlier, one such criterion is the (statistical) fidelity, Φ: usually, periods with $\Phi \geq 12$ reflect some dynamical process that acted upon the data sampled. Physical constraints also could be invoked. For example, since we know that life on Earth does not die out every precession cycle, the precession alone is not a mass-extinctions driver. Likewise, we know that the Earth's magnetic poles do not revert (the geomagnetic field does not change polarity) every precession either, so the precession alone is not a polar-reversals forcer, and so on. These physical constraints say nothing about the precessional resonance since it includes numerous mechanical waves of varying energy levels and destruction potential.

Thus the Earth axial precession's modulation due to annual variation in the Equation of Time (Fig. 3), P=9.34737 My, has been reported previously and repeatedly as a genuine and statistically significant spectral peak. In one of the more extreme examples of confusing means and ends, Martinez and Dera (2015), while aware that "grand orbital cycles" (on the order of My) are just modulations of astronomical cycles, nevertheless declare a ~9-My peak they found in the Jurassic–Early Cretaceous carbon cycle also physically significant. They even proclaim it "an important metronome of the greenhouse climate dynamics" — all based on statistics alone.

Other workers, however, do recognize this modulation as physically unrealistic and proceed on to examine possible non-orbital causes for its presence; see, e.g., Boulila et al. (2012) study of a ~9-My pseudoperiod in Cenozoic carbon isotope record δ13C but who also fail to explain the presence of this pseudoperiod. In an assessment of a recent example from a non-marine tetrapod extinction record (compiled by Rampino et al., 2020) insensitive to the ocean-tidal component as the most significant constituent of systematic noise, Omerbashich (2021) found that most, if not all, the periods claimed previously as mass-extinction cycles are instead astronomical reflections — of Earth's axial precession primarily, i.e., n·p. A mass-extinctions period of 27.5 My extracted using circular and Fourier methods from their compilation and cratering by Rampino and Caldeira (2015) was questioned previously by Meier and Holm–Alwmark (2017) over circular method use and then by Omerbashich (2021) over Fourier method use. As mentioned, (*ibid*.) showed that the 27.5-My period is just a ~3P modulation (the above-discussed tripling in multi-rigid-body resonances, e.g., Prokoph and Puetz, 2015). Previously, Omerbashich (2006, 2007b) had demonstrated from a flawed report by Rohde and Muller (2005) that data manipulations and inapt methods like the Fourier class of spectral analysis techniques can result in a detection of apparently significant periods in marine diversity data as well.

Since P is controlled primarily by the Earth's own orbital (annual) cycle, subsequently driven into secular, centennial, millennial modulations, and on — the total modulation of P extends virtually *ad infinitum*. Thus, the results of the present study are precession rate-independent and readily applicable to any geological interval to Triassic, thanks to the strict nature of paleomagnetism primarily (Berger, 2012), but also beyond since the main find is strictly astronomical in origin. For example, 10P and 100P/7 reflections were first reported by Omerbashich (2006) from the Sepkoski fossils record as 91.3-My and 140.23-My at the ≥99%-significance level. Boulila et al. (2018) confirmed the former of the two periods but also reported a ~9.3-My cyclicity as previously attributed to the long-period Milankovitch band and based on the Cenozoic record. In addition, *(ibid.*) reported a ~250-My megacycle they even attributed to tectonic-causal mechanisms. However, this periodicity also is just an annually-secularly induced reflection 100P/4, Fig. 3. The reason for the latter period to be so persistent (very long) is that it is a crosscycle — amplified and maintained by the 70P' reflection as well, Fig. 5.

Other reports of reflections — sensationally declared to be key or even justified by cryptic galactic causes — fill the literature and are often picked by national and international media in search of a sensation more. In one of the most notorious examples, Boulila (2019) reported the "prominent ~9 and ~36 My" cyclicities in Cenozoic–Cretaceous benthic foraminifera $\delta^{18}O$, but a look at Fig. 3 tells that those are just annually-secularly induced P and its ~4P reflection, respectively. The only reason the often reported ~36-My period appears so persistent and omnipresent across the geological record is that it too is a crosscycle — amplified by 10P' reflection, Fig. 5, as well as the 2π (circular) modulation 360°·p'', where p'' is the Earth eccentricity astronomical period, of ~100 ky.

More of similarly simple mathematical relations are noticed readily among detected periods in paleodata. For example, looking at the nine periods from Table 2 of Omerbashich (2021), the 33.84146-My period is just the 8.48624-My period quadrupled, or 6.67870-My quintupled, while the 11.93548-My period is simply the 7.95129-My period tripled-halved and so on. As mentioned earlier, these subharmonic relations — doubling, halving, and tripling primarily — represent discrete time translation symmetry characteristic of subharmonic resonant response of multi-rigid-body systems to external periodic forcing. For instance, the above-discussed ~3P periods are often reported in the literature, but such reports used mostly inept techniques unable to even distinguish resonances from resonance drivers.

The type of result obtained in the present study calls for an addition to the Milankovitch theory (see, e.g., Berger, 2012) to differentiate among varying power of the Milankovitch — as demonstrated here — resonant cycles. This varying power from one to another resonance wave explains the 100-ky cyclicity problem of that theory — the otherwise inexplicable termination events when astronomical cycles (and, in the sense of resonance theory, their reflections or harmonics) switch in the role of planetary change driver. The latest recorded event of glacial-interglacial switching, the Mid-Pleistocene Transition, occurred ~1 My ago. At that time, the ~40-ky period gave in to the ~100-ky period (Bajo et al., 2020). This event resembles an energy-transfer event across frequencies — the relatively brief period of nonlinear resonance (Rial et al., 2013), here demonstrated to be incessant due to astronomical forcing. The difficulties in assigning physical mechanisms to generate the gain necessary to power the 100 ky cycles led Liu (1992) to examine frequency resonance phenomena among the orbital parameters as an alternative source for the transition, and he concluded that it is the Earth obliquity which modulates a major 100 ky periodicity (Hinnov, 2000). As shown earlier, and supporting that result, the Mid-Pleistocene switch from the obliquity's p' period to a 100-ky period was not a transition from obliquity to eccentricity as the Earth's dominant climatic driver (for which there would be no explanation), but an obliquity-triggered resonant jump from one to another resonant frequency coincidentally near the duration of one eccentricity cycle. The orbitally induced and obliquity-triggered precessional resonance is the most reasonable physics-based, all-explanatory addition to the intellectually already pleasing Milankovitch theory.

The addition to the Milankovitch theory is overdue, so to recognize that that theory is just a special case of the far more complex Earth–Moon–Sun (rather: obliquity–precession–Earth) resonant energy transfer. Similarly, that theory's spatiotemporal domain extends only for as long as a specialized energy band (currently determined by one of the tailing harmonics of the precessional resonance) of the resonant-dynamic energy band that, at present, envelopes and shapes our planet lasts before it dissipates. As it dissipates, it does it only to make room for another precessional-resonant energy band belonging to another resonance harmonic (and so on, until $P_R$) and which will be describable by some other form of the Milankovitch theory — not entirely unlike the current one. The energy supplied thus to the Earth is globally cumulative since the Earth acquired the Moon ~4450 billion years ago, and thus the only known mechanism for a growing Earth proposed by many, e.g., Schmidt and Embleton (1981) and Maxlow (2021). It is also an all-inclusive explanation of the Great Unconformity — as a global resonant shake-off during a TRE (virtually instantaneous ubiquitous erosion of an entire chunk of strata, akin to more frequent decimations of life records in strata creating an impression of periodic mass extinctions). Omerbashich (2020a) proposed that the smaller of the two Mars's moons be reassigned to Earth to create a permanent interference before the Earth reaches the state of energy overload and implodes.

The present study revealed that the concept of time crystal known from quantum physics is also macroscopic at astronomical scales. This synchronization of quantum type in vibrations of dynamic structures at interplanetary and interstellar scales could arise tidally due to a satellite galaxy or universe — a common concept from astrophysics. In the first attempt to link the Milankovitch theory to Earth's internal processes at My time scales, Boulila et al. (2021) erroneously speculated that the most likely phenomenon linking insolation-climate with the Earth's internal processes lies in climatically-resonantly driven mass changes on Earth surface, such as imagined sea level variations due to climate change. In their opinion, the potential link between external and internal processes would thus manifest as interaction and feedback rather than a direct orbital forcing of the solid Earth. However, as showed in the present study computationally and confirmed by exposing Earth's behavior as that of a gigantic macroscopic time crystal, the exact opposite holds instead: direct orbital forcing of the solid Earth, her interior, and the Milankovitch process has been taking place in unison since time immemorial. This cross-scale-verified computational discovery conclusively confirms the genuine nature of the South Atlantic Anomaly of the geomagnetic field low as a precursor of an upcoming reversal, contrary to recent speculations, e.g., by Nilsson and Muscheler (2022).

## 11. Conclusions

Discrete time translation symmetry (seen commonly in paleodata as period multiplication and halving), 2π-phase-shifts, and ¼-lockstep to the forcer — previously thought to arise together only in a subharmonic response of a time crystal to external periodic forcing and to be confined to quantum scales — characterize the here reported global geophysical detection of the mechanism for geomagnetic polarity reversals from the CKGPTS95 timescale calibration as arguably the best record of paleodata available for this type of study. This result has fundamental implications as it directly relates phenomena from quantum physics and macroscopic astrophysics: *the (planetary) time crystal is formed by the many-body entrainment causing precession resonance that, in turn, moderates (geo)magnetic polarity*. In quantum dynamics, such particle entrainment can take the form of collisions, causing Feshbach resonances.

This fundamental property was arrived at by analyzing paleodata periods previously claimed in the literature as standalone-significant cyclicity (indicative of a dynamical process) but which turned out to be just multiples or fractions of other periods appearing in extinctions and other geological records. Thus previously claimed very long periods in paleodata, such as the often reported ~9-My, are primarily modulations of the Earth's axial precession. Together with respective modulations (reflections and harmonics), they arise resonantly under obliquity and other triggers such as varying crustal thickness, mantle flows, and inner-core dynamics — essentially any obstacles in the path of macroscopic Faraday instabilities. To avoid misattributing causality, future analyses of non-astronomical periodicity in paleodata should enforce (i.e., mathematically ignore) precession periods and their annual 2π-phase-shifted modulations and entire precessional resonance ensembles, depending on the band or scale of extraction.

The global approach to treating the currently accepted CKGPTS95 timescale of geomagnetic polarity reversals, viewed as the gauge for analyzing planetary paleodynamics, has confirmed the general absence of lonesome (including extraterrestrial) mechanisms for periodic polarity reversals. At the same time, that approach enabled deducing the 0.02–40-My bands of resonance using the 26.5-My Rampino period as a carrier of cumulative resonant destruction of obstacles. As the orbital energy transfer occurs in these mechanical resonances, their low-frequency part can become transformatively destructive and plate tectonics-catalyzing; this energy transfer constitutes a plausible energy-injecting mechanism for astronomical forcing of climate, albeit in previously underappreciated ways. Namely, the 2π-phase-shifted precessional resonance extracted here in the 1–40 My band potentially represents the scale-invariant energy-transfer mechanism underlying the Milankovitch theory as a special case applicable to the current stage of the Earth-Moon-Sun resonant dynamics.

Sedimentary stratigraphy may therefore record this orbital-to-mechanical resonant energy transfer so that long-period (My) scales of resonant periodicity need not envoke a major extraterrestrial (galactic) or tectonic driver. Instead, the precessional resonance, during rare but cataclysmic Transformative Resonant Events (TREs) characterized by combined destructive interference and angular deflection of its most energetic parts via the 26.5-My Rampino carrier wave, could downward penetrate the inner core, envelope it, and mechanically flip Earth cores, thereby causing geomagnetic polarity reversals. In doing so, while modifying all solids, including the crust, these events may also decimate the geological record containing evidence of evolution, creating the apparent periodicity of mass extinctions and even strata voids like the Great Unconformity. Over geological history, this continuous resonance might have caused geomagnetic polarity excursions while masking sudden cataclysms under apparently gradual processes. Other geophysical theories previously dismissed for lacking mechanism, such as expanding Earth, may now appear more plausible.

It can be speculated that the upcoming geomagnetic polarity flip may manifest as either a mere excursion or a full reversal; in the latter case, an accompanying TRE might occur. Given the thousands of harmonics between consecutive polarity reversals and the millions of minor reflections accumulated via repetitive annual imprinting (the projector effect), it might be plausible to assume that future architecture and urbanism should consider survival under protective habitats, such as underlake and undersea superbunkers independent of topography, until successful space colonization is achieved.

While the present analysis focuses on geophysical mechanisms, the analogy to quantum time-crystal behavior underscores that resonance and periodicity are universal phenomena across scales. Additional research directions arise from the cross-disciplinary aspects of these results. It can be speculated that applications of global geodynamics in quantum dynamics and vice versa may be fruitful. Specifically, the Earth-Moon-Sun triad was shown here to be in a coherent state resembling that of a quantum discrete time crystal. In quantum mechanics, a coherent state is one whose dynamics most closely resemble the oscillatory behavior of a classical harmonic oscillator. However, since quantum coherent states are difficult to maintain under thermal noise, and subscale instabilities cannot project perfectly onto the global system, it can be argued that quantum time crystals may, in fact, derive their observed behavior from macroscopic phenomena and never the other way around.

Consequently, quantum (discrete) time-crystal phenomena, as well as entanglement, tunneling, and other quantum effects, might be considered less surprising when viewed from the planetary scale. Based on the cosmological and astrophysical principle of (always downwardly) continuing, cross-scale projecting of periodic gravitational phenomena and magnetic properties, we can expect that the Earth's resonant system sets the tone for all subscale resonances, including those at quantum scales and below. Accordingly, quantum physics may manifest differently outside the vicinity of Earth—as a highly localized phenomenon of higher-order tidal resonance constrained by the host stellar system. Because the tidal phenomenon is pervasive and general across the known universe, these considerations suggest that terrestrial (now simplistic) estimates of the masses of all fundamental particles—including quarks, leptons, gauge bosons, the Higgs boson, and also properties related to dark matter and dark energy (via accumulated effects of quantum and resonance variations across the universe)—whose mass calculations critically depend on local quantum fluctuations, are influenced or actually explained by higher-order nonlinear geodynamic tidal phenomena rather than solely by universal quantum properties.

**Acknowledgments**

The spectra were computed using LSSA v.5 scientific software based on Vaníček (1969; 1971), http://web.archive.org/web/20220818070617id_/http:/www2.unb.ca/gge/Research/GRL/LSSA/sourceCode.html.

**Statements**

The author declares no conflicts of interest.

**Data access**

All data generated and/or analyzed during the present study are included in this article.